\documentclass[11pt]{article}

\usepackage{microtype} 
\usepackage{booktabs}  
\usepackage{url}  
\usepackage{float}
\usepackage{comment}

\usepackage{amsmath}
\usepackage{amsthm}
\usepackage{graphicx}
\usepackage{adjustbox}

\usepackage{multirow}
\usepackage{acro}
\acsetup{list/template=description}
\DeclareAcronym{RMSD}{
	short = RMSD,
	long = root mean square deviation of atomic positions,
}
\DeclareAcronym{qHTS}{
	short = qHTS,
	long = quantitative high-throughput screening,
}

\DeclareAcronym{LLM}{
	short = LLM,
	long = large language model,
}

\DeclareAcronym{PLM}{
	short = PLM,
	long = protein language model,
}

\DeclareAcronym{MT-DNN}{
	short = MT-DNN,
	long = multi-task deep neural network,
}

\DeclareAcronym{jpg}{
	short = JPEG ,
	sort = jpeg ,
	alt = JPG ,
	long = Joint Photographic Experts Group
}

\DeclareAcronym{ML}{
	short = ML,
	long = machine learning
}

\DeclareAcronym{DL}{
	short = DL,
	long = deep learning
}

\DeclareAcronym{MCC}{
	short = MCC,
	long = Matthews correlation coefficient
}

\DeclareAcronym{Sp}{
	short = Sp,
	long = specificity
}

\DeclareAcronym{Sn}{
	short = Sn,
	long = sensitivity
}

\DeclareAcronym{BA}{
	short = BA,
	long = balanced accuracy
}

\DeclareAcronym{AP}{
	short = AP,
	long = average precision
}

\DeclareAcronym{BEDROC}{
	short = BEDROC,
	long = Boltzmann-enhanced discrimination of receiver operating characteristic
}

\DeclareAcronym{ROC_AUC}{
	short = ROC AUC,
	long = receiver operating characteristic area under curve
}

\DeclareAcronym{PR_AUC}{
	short = PR AUC,
	long = precision recall area under curve
}

\DeclareAcronym{DPR_AUC}{
	short = $\Delta$PR AUC,
	long = $\Delta$ in precision recall area under curve
}

\DeclareAcronym{TPR}{
	short = TPR,
	long = true positive rate
}

\DeclareAcronym{TNR}{
	short = TNR,
	long = true negative rate
}

\DeclareAcronym{FPR}{
	short = FPR,
	long = false positive rate
}

\DeclareAcronym{FNR}{
	short = FNR,
	long = false negative rate
}

\DeclareAcronym{TP}{
  short = TP,
  long  = true positive
}

\DeclareAcronym{FN}{
  short = FN,
  long  = false negative
}

\DeclareAcronym{FP}{
  short = FP,
  long  = false positive
}

\DeclareAcronym{TN}{
  short = TN,
  long  = true negative
}

\DeclareAcronym{RF}{
	short = RF,
	long = random forest
}

\DeclareAcronym{AID}{
	short = AID,
	long = bioassay identifier
}

\DeclareAcronym{HTS}{
	short = HTS,
	long = high-throughput screening
}

\DeclareAcronym{MMP}{
	short = MMP,
	alt = $\Delta\Psi_\text{m}$,
	long = mitochondrial membrane potential 
}

\DeclareAcronym{m-MPI}{
	short = m-MPI,
	long = mitochondrial membrane potential indicator
}

\DeclareAcronym{ECACC}{
	short = ECACC,
	long = European Collection of Authenticated Cell Cultures 
}

\DeclareAcronym{DMEM}{
	short = DMEM,
	long = Dulbecco's modified eagle medium 
}

\DeclareAcronym{FCS}{
	short = FCS,
	long = fetal calf serum
}

\DeclareAcronym{RT}{
	short = RT,
	long = room temperature
}

\DeclareAcronym{FCCP}{
	short = FCCP,
	long = carbonylcyanid-4-(trifluormethoxy)phenylhydrazon,
}

\DeclareAcronym{DMSO}{
	short = DMSO,
	long = dimethyl sulfoxide,
}

\DeclareAcronym{ddH2O}{
	short = \ch{ddH2O},
	long = double destilled water,
	sort={ddH2O},
}

\DeclareAcronym{PBS}{
	short = PBS,
	long = phosphate-buffered saline,
}

\DeclareAcronym{EC50}{
	short = EC\textsubscript{50},
	long = half maximal effective concentration,
}

\DeclareAcronym{AI}{
	short = AI,
	long = artificial intelligence,
}

\DeclareAcronym{DTI}{
	short = DTI,
	long = drug--target interaction,
}

\DeclareAcronym{DDI}{
	short = DDI,
	long = drug--drug interaction,
}

\DeclareAcronym{DNA}{
	short = DNA,
	long = deoxyribonucleic acid,
}

\DeclareAcronym{ECFP}{
	short = ECFP,
	long = extended-connectivity fingerprint,
}

\DeclareAcronym{FCFP}{
	short = FCFP,
	long = functional-class fingerprint,
}

\DeclareAcronym{MAP4}{
	short = MAP4,
	long = MinHashed atom-pair fingerprint,
}

\DeclareAcronym{SVM}{
	short = SVM,
	long = support vector machine,
}

\DeclareAcronym{DNN}{
	short = DNN,
	long = deep neural network,
}

\DeclareAcronym{GCNN}{
	short = GCNN,
	long = graph convolutional neural networks,
}

\DeclareAcronym{GHS}{
	short = GHS,
	long = globally harmonized system of classification and labelling of chemicals,
}

\DeclareAcronym{SMILES}{
	short = SMILES,
	long = simplified molecular-input line-entry system,
}

\DeclareAcronym{CLI}{
	short = CLI,
	long = command-line interpreter,
}

\DeclareAcronym{GUI}{
	short = GUI,
	long = graphical user interface,
}

\DeclareAcronym{ATP}{
	short = ATP,
	long = adenosine triphosphate,
}

\DeclareAcronym{AMP}{
	short = AMP,
	long = adenosine monophosphate,
}

\DeclareAcronym{PPi}{
	short = PP\textsubscript{i},
	long = pyrophosphate,
	sort={PPi}
}

\DeclareAcronym{Lu}{
	short = \ch{LH2},
	long = luciferin,
	sort={LH2}
}

\DeclareAcronym{oLu}{
	short = \ch{oxy-L},
	long = oxy-luciferin,
	sort={oxylu}
}

\DeclareAcronym{SMOTE}{
	short = SMOTE,
	long = synthetic minority over-sampling technique,
}

\DeclareAcronym{SHAP}{
	short = SHAP,
	long = Shapley additive explanation,
}

\DeclareAcronym{CPU}{
	short = CPU,
	long = central processing unit,
}

\DeclareAcronym{RAM}{
	short = RAM,
	long = random-access memory,
}

\DeclareAcronym{GPU}{
	short = GPU,
	long = graphics processing unit,
}

\DeclareAcronym{CAS}{
	short = CAS,
	long = Chemical Abstracts Service,
}

\DeclareAcronym{UMAP}{
	short = UMAP,
	long = uniform manifold approximation and projection,
}

\DeclareAcronym{Tox21}{
	short = Tox21,
	long = Toxicology in the 21\textsuperscript{st} Century,
}

\DeclareAcronym{GOSS}{
	short = GOSS,
	long = gradient-based one-side sampling,
}

\DeclareAcronym{QSAR}{
	short = QSAR,
	long = quantitative structure--activity relationship,
}

\DeclareAcronym{EFB}{
	short = EFB,
	long = exclusive feature bundling,
}

\DeclareAcronym{MTT}{
	short = MTT,
	long = \iupac{3-(4, 5-dimethylthiazolyl-2)-2,5-diphenyltetrazolium bromide},
}

\DeclareAcronym{SSL}{
	short = SSL,
	long = self-supervised learning,
}

\DeclareAcronym{GBM}{
	short = GBM,
	long = gradient boosting machine,
}

\DeclareAcronym{MLP}{
	short = MLP,
	long = multilayer perceptron,
}

\DeclareAcronym{API}{
	short = API,
	long = application programming interface,
}

\DeclareAcronym{CHOP}{
  short = CHOP,
  long  = C/EBP homologous protein
}
\DeclareAcronym{UPR}{
  short = UPR,
  long  = unfolded protein response
}
\DeclareAcronym{ER}{
  short = ER,
  long  = endoplasmic reticulum
}

\DeclareAcronym{PN}{
  short = PN,
  long  = prototypical network
}

\DeclareAcronym{FH}{
  short = FH,
  long  = frequent hitters
}

\DeclareAcronym{LSA}{
  short = LSA,
  long  = latent semantic analysis
}

\DeclareAcronym{ADMET}{
  short = ADMET,
  long  = {absorption, distribution, metabolism, excretion and toxicity}
}

\DeclareAcronym{GNN-ST}{
  short = GNN-ST,
  long  = single-task graph neural network
}

\DeclareAcronym{GNN-MT}{
  short = GNN-MT,
  long  = multi-task graph neural network
}

\DeclareAcronym{GNN-MAML}{
  short = GNN-MAML,
  long  = model-agnostic meta-learning graph neural network
}

\DeclareAcronym{MAT}{
  short = MAT,
  long  = molecule attention transformer
}

\DeclareAcronym{1D}{
    short = 1D,
    long  = one-dimensional 
}

\DeclareAcronym{2D}{
    short = 2D,
    long  = two-dimensional 
}

\DeclareAcronym{3D}{
    short = 3D,
    long  = three-dimensional 
}

\DeclareAcronym{ITC}{
    short = ITC,
    long  = isothermal titration calorimetry
}

\DeclareAcronym{LA}{
    short = LA,
    long  = lipoic acid
}
\usepackage{siunitx}
\usepackage[preprint]{automl}
\usepackage[
    numbers, 
    sort&compress, 
    super,
]{natbib}
\usepackage{csquotes}
\usepackage{hyperref}
\usepackage{cleveref}

\renewcommand{\cref}[1]{\Cref{#1}}
\Crefname{equation}{Eq.}{Eqs.}
\Crefname{figure}{Fig.}{Figs.}
\Crefname{table}{Tab.}{Tabs.}
\Crefname{tabular}{Tab.}{Tabs.}
\crefname{equation}{Eq.}{Eqs.}
\crefname{figure}{Fig.}{Figs.}
\crefname{table}{Tab.}{Tabs.}
\crefname{tabular}{Tab.}{Tabs.}

\title{Barlow Twins Deep Neural Network for Advanced 1D Drug--Target Interaction Prediction}

\author[1]{\nameemail{Maximilian G. Schuh}{m.schuh@tum.de}}
\author[1]{\nameemail{Davide Boldini}{davide.boldini@tum.de}}
\author[2]{\nameemail{Annkathrin I. Bohne}{a.bohne@tum.de}}
\author[1]{\nameemail{Stephan A. Sieber}{stephan.sieber@tum.de}}

\affil[1]{Technical University of Munich, TUM School of Natural Sciences, Department of Bioscience, Center for Functional Protein Assemblies (CPA), Chair of Organic Chemistry II, 85748 Garching bei München, Germany}
\affil[2]{Technical University of Munich, TUM School of Natural Sciences, Department of Bioscience, Center for Functional Protein Assemblies (CPA), Chair of Biochemistry, 85748 Garching bei München, Germany}

\hypersetup{%
  pdfauthor={Maximilian G. Schuh, Davide Boldini, Annkathrin I. Bohne, Stephan A. Sieber}, 
  pdftitle={Barlow Twins Deep Neural Network for Advanced 1D Drug-Target Interaction Prediction},
  pdfsubject={},
  pdfkeywords={Drug-Target Interaction, Machine Learning, Deep Learning, Drug Discovery}
}

\newcommand{\BTDTI}{\textsc{BarlowDTI}~}
\newcommand{\BTDTIXXL}{\textsc{BarlowDTI}\textsubscript{XXL}~}
\newcommand{\BTDTIwospace}{\textsc{BarlowDTI}}
\newcommand{\BTDTIXXLwospace}{\textsc{BarlowDTI}\textsubscript{XXL}}

\begin{document}

\maketitle

\begin{abstract}
Accurate prediction of \aclp{DTI} is critical for advancing drug discovery.
By reducing time and cost, \acl{ML} and \acl{DL} can accelerate this laborious discovery process.
In a novel approach, \BTDTIwospace, we utilise the powerful Barlow Twins architecture for feature-extraction while considering the structure of the target protein.
Our method achieves state-of-the-art predictive performance against multiple established benchmarks using only \acl{1D} input.
The use of \acl{GBM} as the underlying predictor ensures fast and efficient predictions without the need for substantial computational resources.
We also investigate how the model reaches its decision based on individual training samples.
By comparing co-crystal structures, we find that \BTDTI effectively exploits catalytically active and stabilising residues, highlighting the model's ability to generalise from \acl{1D} input data.
In addition, we further benchmark new baselines against existing methods.
Together, these innovations improve the efficiency and effectiveness of \acl{DTI} predictions, providing robust tools for accelerating drug development and deepening the understanding of molecular interactions.
Therefore, we provide an easy-to-use web interface that can be freely accessed at \url{https://www.bio.nat.tum.de/oc2/barlowdti}.
\end{abstract}

\acresetall

\section{Introduction}

Studying \acp{DTI} is crucial for understanding the biochemical mechanisms that govern how molecules interact with proteins.\cite{rang2011rang}
Key challenges in drug discovery are the identification of proteins that can be used as targets for the treatment of diseases.\cite{strittmatter2014overcoming}
To achieve the desired therapeutic effects, the discovery of molecules that interact with and activate or inhibit target proteins is essential.\cite{hughes2011principles,blundell2006structural,tautermann2020current}

Recent advances in computational methods have transformed the drug discovery landscape, providing robust tools for cost-effective exploration of the chemical space.
These \textit{in silico} approaches facilitate the prediction and analysis of \acp{DTI}, aiding in the identification of potential drug candidates and their corresponding protein targets.\cite{agu2023molecular,bender2021practical,hollingsworth2018molecular,karplus1990molecular,dhakal2022artificial,you2022artificial}
The use of computational techniques allows researchers to gain a comprehensive understanding of the molecular mechanisms underlying \acp{DTI}, thereby accelerating the drug discovery process and minimising reliance on traditional, resource-intensive experimental methods.\cite{kitchen2004docking,hopkins2009predicting}
Different methods have been used to understand how drugs interact with target proteins. 
These methods are grouped into three main categories: structure-agnostic, structure-based and complex-based.

Structure-agnostic approaches use \ac{1D} representations like molecule \ac{SMILES} and protein amino acid sequences, or \ac{2D} representations like graphs and predicted contact maps.\cite{chen2020transformercpi,jiang2020drug,koh2024physicochemical,lee2024dlmdti}
These methods are cost-effective and and sufficiently accurate compared to experimental or \textit{in silico} structure prediction,\cite{jiang2022sequencebased} as they are independent of the protein's structure when predicting effects. 

Structure-based approaches require \ac{3D} protein structures and \ac{1D} or \ac{2D} molecular inputs.
\ac{3D} structures are usually derived from experimental data, although computational predictions are increasingly employed.\cite{ahdritz2024openfolda,abramson2024accurate,krishna2024generalized,trott2010autodock,corso2023diffdock}
These methods have great potential but can be unreliable. 
They depend on accurate \ac{3D} protein structures and may be limited in their ability to generalise beyond experimentally observed \acp{DTI}.\cite{he2023alphafold2}
Due to the complexity of the experimental setup, \ac{3D} protein structures can be difficult to obtain. 
In addition, models often overlook the fact that proteins are not rigid structures, but are generally in motion, e.g., ligand binding induces a conformational change.\cite{abramson2024accurate,trott2010autodock,corso2023diffdock}

Finally, complex-based approaches require protein-ligand co-crystal structures, which additionally require \ac{3D} information, as well as protein interaction information about the ligand.\cite{li2021structureaware}
For this reason, complex-based approaches can provide a more detailed insight into the interactions, but they are by far the most difficult to obtain data for.

Considering these different approaches, we designed \BTDTI as a fully data-driven, sequence-based approach that relies on \ac{SMILES} and amino acid sequences as the most accessible data, avoiding costly and time-consuming experimental data such as crystal structures.
Additionally, we use a specialised bilingual \ac{PLM} to embed the \ac{1D} amino acid sequence, which uses a \ac{3D}-alignment method that results in a \enquote{structure-sequence} representation.\cite{heinzinger2024bilingual,vankempen2024fast}
This approach makes \BTDTI input data structure-agnostic, yet benefits from \enquote{structure-sequence} \ac{PLM} embeddings.
Unlike most other methods, we have developed a system that uses a hybrid \enquote{best of both worlds} \ac{ML} and \ac{DL} approach to improve \ac{DTI} prediction performance in low data regimes where training data is limited.\cite{chen2016xgboost,schuh2024synergizing}
We have found that \ac{DL} architectures such as Barlow Twins\cite{zbontar2021barlow,barlow1961possible} are excellent at learning features\cite{schuh2024synergizing} that can then be used for \ac{GBM} training to achieve state-of-the-art performance, as the size of datasets is usually too small to reliably train a \ac{DL} model that will perform competitively.

\begin{figure}[!htb]
    \centering
    \includegraphics[height=0.6\textheight]{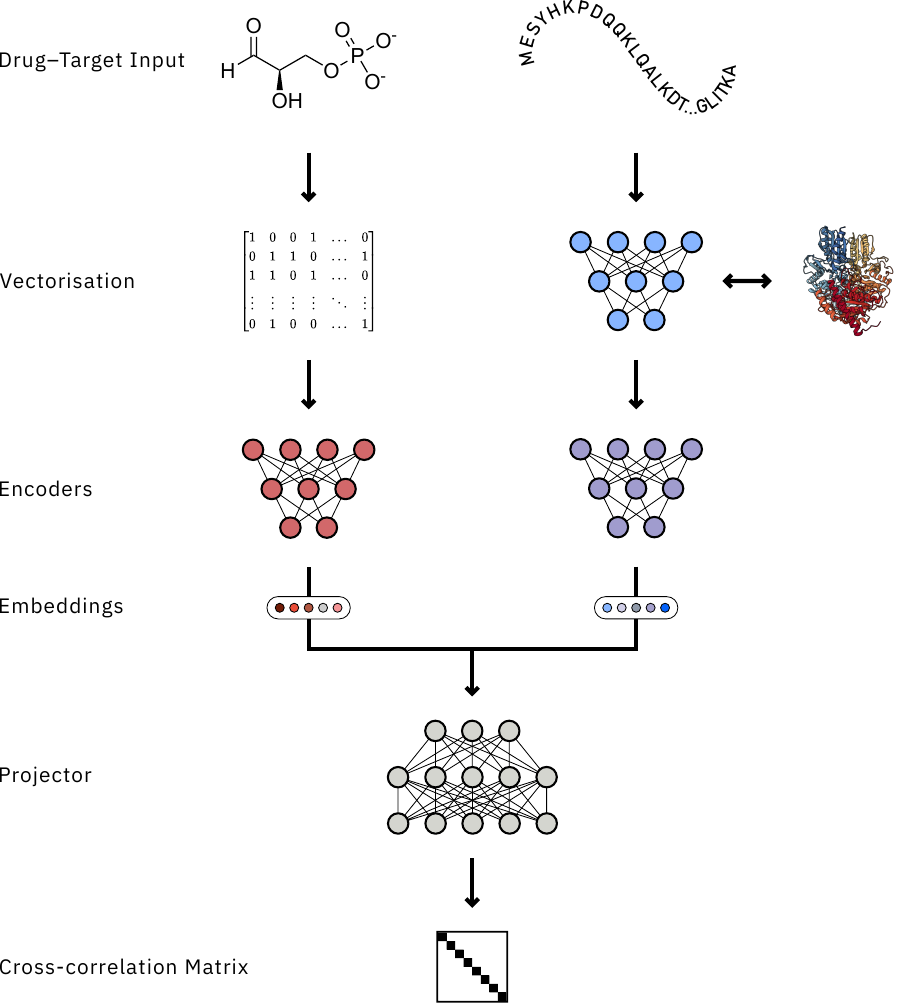}
    \caption{\textbf{\BTDTI architecture.} Drug and target serve as \acs{1D} input, where they are processed and converted into vectors. Molecules are provided as \acs{SMILES} and converted to \acs{ECFP}. On the other hand, the primary amino acid sequence is vectorised using a bilingual \acs{3D} structure-aware \acs{PLM}. The Barlow Twins architecture learns to understand \acsp{DTI}. The objective function forces both representations of the \ac{DTI} to be as close as possible to the unity matrix. Finally, this \acs{DL} model is used as a feature-extractor and a \acs{GBM} is trained on the embeddings and the interaction label. The \acs{GBM} is then used as the predictor.}
    \label{fig:architecture}
\end{figure}

To overcome the limitation of data scarcity, we built \BTDTIXXLwospace, which is trained on millions of curated \ac{DTI} pairs,\cite{golts2024large} to apply the model to real-world examples, as we have done in case studies.
Here, \BTDTIXXL captures the correlation between experimentally determined affinities and the predicted likelihood of interaction, proving our approach useful in drug discovery settings.
By comparing co-crystal biochemical structures and their active sites, we also investigate and explain how \BTDTIXXL arrives at its decision.
We conduct our investigation by employing an influence method and adapting it in a novel way to identify the most important training \acp{DTI}.\cite{brophy2022instancebased}
This work culminates in a freely available web interface that takes \ac{1D} input of molecule and protein information and predicts the likelihood of interaction. 

\section{Results and Discussion}

\paragraph{\BTDTI design}
We propose a novel method for predicting \acp{DTI} using \ac{SMILES} notations, primary amino acid sequences, both \ac{1D}, and annotated interaction properties.
\BTDTI relies on a several key components, visualised in \cref{fig:architecture}:

\begin{enumerate}
    \item Firstly, the input needs to be vectorised. This is achieved by converting \ac{SMILES} to an \ac{ECFP}.
    The amino acid sequences are processed by a \ac{PLM} that uses both modalities, combining \ac{1D} protein sequences and \ac{3D} protein structure.\cite{heinzinger2024bilingual}
    \item Secondly, we teach the \ac{SSL} based Barlow Twins model interaction of molecule and protein without considering labels.\cite{zbontar2021barlow,barlow1961possible}
    The objective function implements invariance of both representation of one interaction while ensuring non-redundancy of the features.\cite{zbontar2021barlow,barlow1961possible}
    \item Finally, \BTDTI takes a combination of embeddings generated by the encoders from the Barlow Twins \ac{DL} model and uses them as features to train a \ac{GBM} based on the interaction annotations.\cite{chen2016xgboost}
    This approach exploits two key strengths: it uses \ac{DL} to refine representations, and it leverages the power of \ac{ML} in scenarios with limited data. 
    This is particularly relevant for current \ac{DTI} benchmarks/datasets, where only around \num{50000} annotated pairs are publicly available.\cite{biosnapnets,liu2007bindingdb,davis2011comprehensive,knox2024drugbank}
    Consequently, we propose \BTDTIXXL which is trained on more than \num{3600000} curated \ac{DTI} pairs, additionally sourced from PubChem and ChEMBL,\cite{kim2023pubchem,mendez2019chembl} to obtain generalisability in real-world scenarios.\cite{golts2024large}
\end{enumerate}

\paragraph{Benchmark selection}
We selected a comprehensive set of literature-based benchmarks to evaluate the performance of \BTDTI against several leading methods.
The benchmarks considered in this study are derived from several key sources. 
These sources include biomedical networks,\cite{biosnapnets} the US patent database,\cite{liu2007bindingdb} and data detailing the interactions of 72 kinase inhibitors with 442 kinases, representing over \SI{80}{\percent} of the human catalytic protein kinome.\cite{davis2011comprehensive} 
These datasets provide \acp{DTI} as pairs of molecules and amino acid sequences, each coupled to an interaction annotation.

To ensure a fair comparison, \BTDTI was retrained across all benchmarks. 
Finally, we assessed the model's performance in a binary classification setting, where the task is to distinguish between interacting and non-interacting drug--target pairs:

\begin{itemize}
    \item We compared \BTDTI with a total of seven established \ac{DTI} models: the model by \citeauthor{kang2022finetuning}, MolTrans,\cite{huang2021moltrans} DLM-DTI,\cite{lee2024dlmdti} ConPLex,\cite{singh2023contrastive} DrugBAN,\cite{bai2023interpretable} PSICHIC,\cite{koh2024physicochemical} and STAMP-DTI.\cite{wang2022structureaware}
    For instance, \citeauthor{kang2022finetuning} fine-tuned a \ac{LLM} based on amino acid sequences.\cite{kang2022finetuning}
    MolTrans uses an efficient transformer architecture to increase the scalability of the model.\cite{huang2021moltrans}
    DLM-DTI introduced a dual language model approach combined with hint-based learning to improve prediction accuracy.\cite{lee2024dlmdti} 
    ConPLex leveraged contrastive learning to better understand \acp{DTI},\cite{singh2023contrastive} while DrugBAN focused on interpretable attention mechanisms that provide insights into the interaction process.\cite{bai2023interpretable}
    PSICHIC utilised physicochemical properties to predict interactions more accurately,\cite{koh2024physicochemical} and STAMP-DTI incorporated structure-aware, multi-modal learning to enhance its predictive capabilities.\cite{wang2022structureaware}
    Overall, we evaluated our architecture against the various model implementations.
    These models -- structure-agnostic, structure-based or complex-based -- have demonstrated state-of-the-art performance in benchmarks.
    
    \item This comparison is performed on a total of four datasets with twelve literature-proposed splits: 4 $\times$ BioSNAP,\cite{biosnapnets,kang2022finetuning,koh2024physicochemical} 4 $\times$ BindingDB,\cite{liu2007bindingdb,kang2022finetuning,koh2024physicochemical} 1 $\times$ DAVIS\cite{davis2011comprehensive,kang2022finetuning} and 3 $\times$ Human.\cite{huang2021moltrans,koh2024physicochemical}
    Our aim is to investigate the behaviour of different methods in diverse splitting scenarios, where a whole dataset is split into model training, validation, and evaluation subsets.
    These predefined splits help us to assess how well models generalise under challenging evaluation conditions, for example where either the drug or the target has not been seen before, thus providing insight into their real-world applicability.
    
    \item In addition, we investigated the addition of a more rigorous model baseline. The \ac{GBM} XGBoost is known to be one of the best models, e.g. in \ac{QSAR} tasks, often outperforming \ac{DL}-based approaches.\cite{wu2021we,sheridan2016extreme,asselman2023enhancing}
\end{itemize}

\begin{figure}[!htbp]
    \centering
    \includegraphics[width=\textwidth]{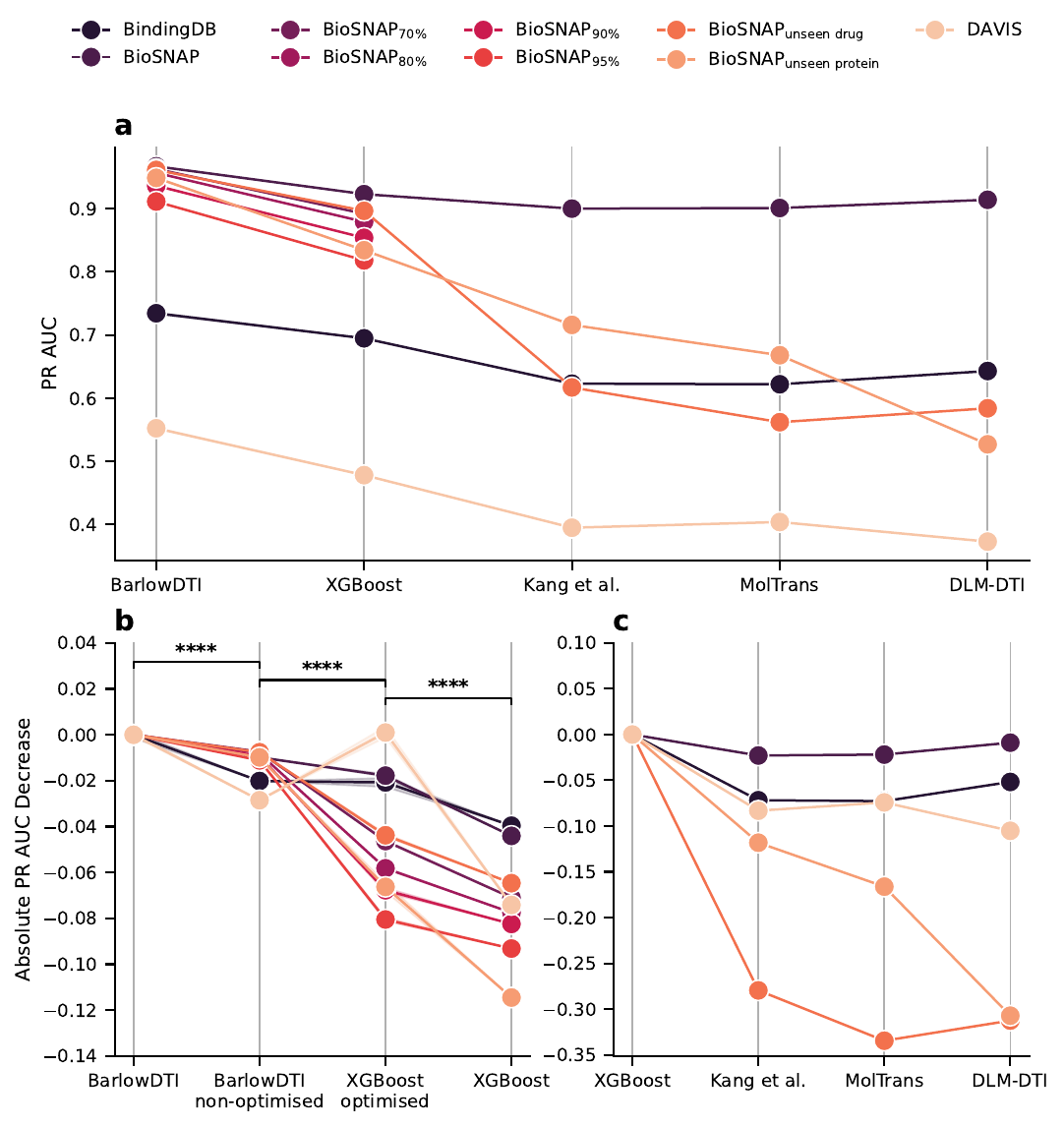}
    \caption{\textbf{A comparison of the performance of methods established in the literature.} 
    \textbf{a}) The state-of-the-art performance of \BTDTI in terms of \ac{PR_AUC} was visualised in comparison to other models (for metrics and their statistics refer to \cref{tab:metrics}). 
    \textbf{b}) The change in performance was examined as key elements of the \BTDTI architecture were incrementally removed. 
    \textbf{c}) The newly introduced model baseline, XGBoost, was compared with other established methods. A per dataset and split difference in \ac{PR_AUC} was calculated based on \BTDTI (in \textbf{b}) performance or the baseline model (in \textbf{c}). The overall change was investigated for statistical significance (****$p < 0.0001$, two-sided Welch's $t$-test,\cite{welch1947generalization,virtanen2020scipy} with Benjamini-Hochberg\cite{benjamini1995controlling} multiple testing correction).}
    \label{fig:absolute_performance_decrease}
\end{figure}

\paragraph{\BTDTI shows state-of-the-art performance in predicting \acp{DTI}}
We assessed the performance of \BTDTI in binary classification across four distinct datasets, each employing different data splitting procedures. 
For each dataset, we predicted whether drug--target pairs in the predefined test subset interact or not. 
We then statistically evaluated these predictions by comparing them to the actual outcomes provided in the benchmark test set, using the metrics \ac{ROC_AUC} and \ac{PR_AUC}.
Overall, \BTDTI significantly outperforms all other models in \cref{fig:absolute_performance_decrease}a and \cref{tab:metrics,tab:p_metrics}.
Looking at BioSNAP, we improve \SI{6}{\percent} over the leading method DLM-DTI in terms of \ac{PR_AUC}.
Furthermore, as shown in \cref{tab:metrics_nature} \BTDTI again outperforms the PSICHIC method with a \SI{7}{\percent} \ac{PR_AUC} improvement independent of the split.

\begin{table}[!htbp]
    \centering
    \caption{\textbf{Benchmarking \BTDTI against other models using \citeauthor{kang2022finetuning} splits.}\cite{kang2022finetuning} Performance was evaluated against three established benchmarks, and the mean and standard deviation of the performance of five replicates are presented. Results per benchmark that are both the best and statistically significant (Two-sided Welch's $t$-test,\cite{welch1947generalization,virtanen2020scipy} $\alpha = 0.001$ with Benjamini-Hochberg\cite{benjamini1995controlling} multiple test correction) are highlighted in bold.}
    \label{tab:metrics}
    \sisetup{
		table-number-alignment = left,
		table-alignment-mode = none,
	}
    \begin{tabular}{llSS}
        \toprule
        \textbf{Dataset} & \textbf{Model} & \text{\textbf{\acs{ROC_AUC}}} & \text{\textbf{\acs{PR_AUC}}} \\
        \midrule
        \multirow{6}{*}{BioSNAP} & \BTDTI & \bfseries 0.9599 (4) & \bfseries 0.9670 (4) \\
              & XGBoost & 0.9142 & 0.9229 \\
              & MolTrans\cite{huang2021moltrans} & 0.895 (002) & 0.901 (004) \\
              & \citeauthor{kang2022finetuning} & 0.914 (006) & 0.900 (007) \\
              & DLM-DTI\cite{lee2024dlmdti} & 0.914 (003) & 0.914 (006) \\
              & ConPLex\cite{singh2023contrastive} & \text{--} & 0.897 (001) \\
        \cline{1-4}
        \multirow{6}{*}{BindingDB} & \BTDTI & \bfseries 0.9364 (3) & \bfseries 0.7344 (18) \\
              & XGBoost & 0.9261 & 0.6948 \\
              & MolTrans\cite{huang2021moltrans} & 0.914 (001) & 0.622 (007) \\
              & \citeauthor{kang2022finetuning} & 0.922 (001) & 0.623 (010) \\
              & DLM-DTI\cite{lee2024dlmdti} & 0.912 (004) & 0.643 (006) \\
              & ConPLex\cite{singh2023contrastive} & \text{--} & 0.628 (012) \\
        \cline{1-4}
        \multirow{6}{*}{DAVIS} & \BTDTI & \bfseries 0.9480 (8) & \bfseries 0.5524 (11) \\
              & XGBoost & 0.9285 & 0.4782 \\
              & MolTrans\cite{huang2021moltrans} & 0.907 (002) & 0.404 (016) \\
              & \citeauthor{kang2022finetuning} & 0.920 (002) & 0.395 (007) \\
              & DLM-DTI\cite{lee2024dlmdti} & 0.895 (003) & 0.373 (017) \\
              & ConPLex\cite{singh2023contrastive} & \text{--} & 0.458 (016) \\
        \bottomrule
    \end{tabular}
\end{table}

When switching to BindingDB, \BTDTI significantly outperforms DLM-DTI in terms of \ac{PR_AUC} with a \SI{>14}{\percent} improvement (\cref{tab:metrics}).
Investigating the BindingDB splits shows that \BTDTI outperforms all existing methods when looking at unseen ligands, matches the \ac{ROC_AUC} performance of DrugBAN in the random setting and becomes second best in the unseen protein split (\cref{tab:metrics_nature}).
Overall, \BTDTI performs best in two out of four splits in this benchmark.

\begin{table}[!htbp]
\centering
\caption{\textbf{Benchmarking \BTDTI against other models using \citeauthor{koh2024physicochemical} splits.}\cite{koh2024physicochemical} Performance was evaluated against three established benchmarks, and the mean of the \BTDTI performance of five replicates are presented. All other metrics are taken from \citeauthor{koh2024physicochemical}. Best result per benchmark and split is highlighted in bold. (\citeauthor{koh2024physicochemical} does not present replicates or sample-correlated predictions.\cite{koh2024physicochemical})}
\label{tab:metrics_nature}
\begin{adjustbox}{totalheight=.8\textheight}
\begin{tabular}{lllSS}
\toprule
\textbf{Dataset} & \textbf{Split} & \textbf{Model} & \textbf{\acs{ROC_AUC}} & \textbf{\acs{PR_AUC}} \\
\midrule
\multirow{15}{*}{BioSNAP} 
& \multirow{5}{*}{Unseen protein} & \BTDTI & \bfseries 0.9572 & \bfseries 0.9679 \\
& & DrugBAN\cite{bai2023interpretable,koh2024physicochemical} & 0.7327 & 0.7971 \\
& & PSICHIC\cite{koh2024physicochemical} & 0.8819 & 0.9071 \\
& & STAMP-DPI\cite{wang2022structureaware,koh2024physicochemical} & 0.8372 & 0.8738 \\
& & XGBoost & 0.8506 & 0.8794 \\
\cmidrule{2-5}
& \multirow{5}{*}{Random split} & \BTDTI & \bfseries 0.9718 & \bfseries 0.9755 \\
& & DrugBAN\cite{bai2023interpretable,koh2024physicochemical} & 0.9089 & 0.9159 \\
& & PSICHIC\cite{koh2024physicochemical} & 0.9246 & 0.9226 \\
& & STAMP-DPI\cite{wang2022structureaware,koh2024physicochemical} & 0.8993 & 0.9056 \\
& & XGBoost & 0.9146 & 0.9242 \\
\cmidrule{2-5}
& \multirow{5}{*}{Unseen ligand} & \BTDTI & \bfseries 0.9666 & \bfseries 0.9706 \\
& & DrugBAN\cite{bai2023interpretable,koh2024physicochemical} & 0.8775 & 0.8843 \\
& & PSICHIC\cite{koh2024physicochemical} & 0.9019 & 0.9030 \\
& & STAMP-DPI\cite{wang2022structureaware,koh2024physicochemical} & 0.8902 & 0.8915 \\
& & XGBoost & 0.8909 & 0.9026 \\
\midrule
\multirow{15}{*}{BindingDB} 
& \multirow{5}{*}{Unseen protein} & \BTDTI & 0.6939 & 0.5791 \\
& & DrugBAN\cite{bai2023interpretable,koh2024physicochemical} & 0.6523 & 0.5295 \\
& & PSICHIC\cite{koh2024physicochemical} & \bfseries 0.7537 & \bfseries 0.6241 \\
& & STAMP-DPI\cite{wang2022structureaware,koh2024physicochemical} & 0.6828 & 0.5735 \\
& & XGBoost & 0.6460 & 0.5233 \\
\cmidrule{2-5}
& \multirow{5}{*}{Random split} & \BTDTI & \bfseries 0.9640 & 0.9513 \\
& & DrugBAN\cite{bai2023interpretable,koh2024physicochemical} & \bfseries 0.9640 & \bfseries 0.9539 \\
& & PSICHIC\cite{koh2024physicochemical} & 0.9503 & 0.9280 \\
& & STAMP-DPI\cite{wang2022structureaware,koh2024physicochemical} & 0.9318 & 0.9085 \\
& & XGBoost & 0.9582 & 0.9462 \\
\cmidrule{2-5}
& \multirow{5}{*}{Unseen ligand} & \BTDTI & \bfseries 0.9456 & \bfseries 0.9263 \\
& & DrugBAN\cite{bai2023interpretable,koh2024physicochemical} & 0.9409 & 0.9188 \\
& & PSICHIC\cite{koh2024physicochemical} & 0.9264 & 0.8975 \\
& & STAMP-DPI\cite{wang2022structureaware,koh2024physicochemical} & 0.9027 & 0.8683 \\
& & XGBoost & 0.9374 & 0.9141 \\
\midrule
\multirow{15}{*}{Human} 
& \multirow{5}{*}{Unseen protein} & \BTDTI & \bfseries 0.9630 & \bfseries 0.9693 \\
& & DrugBAN\cite{bai2023interpretable,koh2024physicochemical} & 0.9298 & 0.9417 \\
& & PSICHIC\cite{koh2024physicochemical} & 0.9503 & 0.9595 \\
& & STAMP-DPI\cite{wang2022structureaware,koh2024physicochemical} & 0.8563 & 0.8748 \\
& & XGBoost & 0.8961 & 0.9171 \\
\cmidrule{2-5}
& \multirow{5}{*}{Random split} & \BTDTI & \bfseries 0.9917 & \bfseries 0.9905 \\
& & DrugBAN\cite{bai2023interpretable,koh2024physicochemical} & 0.9841 & 0.9753 \\
& & PSICHIC\cite{koh2024physicochemical} & 0.9861 & 0.9840 \\
& & STAMP-DPI\cite{wang2022structureaware,koh2024physicochemical} & 0.9659 & 0.9582 \\
& & XGBoost & 0.9813 & 0.9782 \\
\cmidrule{2-5}
& \multirow{5}{*}{Unseen ligand} & \BTDTI & 0.9346 & 0.9348 \\
& & DrugBAN\cite{bai2023interpretable,koh2024physicochemical} & 0.9459 & \bfseries 0.9387 \\
& & PSICHIC\cite{koh2024physicochemical} & \bfseries 0.9500 & 0.9371 \\
& & STAMP-DPI\cite{wang2022structureaware,koh2024physicochemical} & 0.9156 & 0.8980 \\
& & XGBoost & 0.9391 & 0.9337 \\
\bottomrule
\end{tabular}
\end{adjustbox}
\end{table}

\BTDTI once again outperforms all of the established approaches when looking at the DAVIS benchmark, with a \SI{21}{\percent} improvement over the leading ConPLex model in terms of \ac{PR_AUC} (\cref{tab:metrics}).

Lastly, we evaluated the performance on the Human benchmark. 
\BTDTI shows the best performance when looking at the unseen protein split as well as the random split (\Cref{tab:metrics_nature}).
PSICHIC comes first in the unseen ligand setting, when looking at \ac{ROC_AUC}, while DrugBAN is best in \ac{PR_AUC}.
In summary, \BTDTI outperforms all other models in two out of three splits.

We looked at the architecture and its components, removing one at a time and measuring the effect on performance to investigate why \BTDTI outperforms other methods in various benchmarks.

\paragraph{Unravelling the performance contributions of the \BTDTI architecture}
To investigate the impact of each element of the \BTDTI architecture, we removed them one at a time.
We have done this across all baselines and splits with the following ablations:

\begin{enumerate}
    \item We removed the hyperparameter optimisation step of the \BTDTI classifier.  
    \item From the first removal, we replaced the Barlow Twins architecture entirely and instead concatenate \acp{ECFP} and \ac{PLM} embeddings for training. We kept the hyperparameter optimisation procedure as in \BTDTIwospace.
    \item Finally, we removed the hyperparameter optimisation procedure from the previous ablation, analogous to the first modification. 
\end{enumerate}

We observe a significant decline in performance, as illustrated in \cref{fig:absolute_performance_decrease}b and \cref{tab:p_ablation} for the initial ablation, emphasising the crucial role of hyperparameter optimisation for achieving optimal model performance. 

The second ablation also indicates a significant reduction in performance. 
This is likely attributed to the \ac{DL} architecture based on the \ac{SSL} Barlow Twins model, which effectively learns embeddings to describe \acp{DTI}. 
The Barlow Twins objective promotes orthogonality between drug and target modalities while ensuring the non-redundancy of both, thus preventing informational collapse. 
As a result, this leads to an overall state-of-the-art predictive performance.

The final ablation shows a further decline in performance, consistent with the results of the initial ablation experiment. 

In summary, the sustained reduction in performance of our ablation experiments demonstrates that each component of our \BTDTI pipeline is needed to maximise performance. 
This architecture integrates the \enquote{best of both worlds}: \ac{DL} and \ac{GBM} to enhance predictive performance.
Compared to other pure \ac{ML}- or \ac{DL}-based approaches, we can demonstrate a performance boost.
In particular, the use of a state-of-the-art \ac{PLM}\cite{heinzinger2024bilingual} could offer an advantage over other methods.
Other \ac{PLM} variants are ProtTrans T5\cite{elnaggar2022prottrans} in ConPLex\cite{singh2023contrastive} and ProtBERT proposed by \citeauthor{kang2022finetuning} also used in DLM-DTI.\cite{kang2022finetuning}
The structural awareness of \BTDTI added by the inclusion of \ac{3D}-alignment in ProstT5\cite{heinzinger2024bilingual} hints towards better generalisation capabilities, yielding increased performance.

\subparagraph{Choosing baseline models}
Selecting an appropriate baseline model is critical to effectively comparing different \ac{ML} and \ac{DL} techniques. 
Robust baselines are the basis for meaningful comparisons and highlight improvements from new methods.
Without appropriate baselines, it becomes difficult to determine whether new approaches are truly advancing the field.

Current leading \ac{DTI} models predominantly use \ac{DL} methods and are often evaluated against simple baseline models such as logistic regression, ridge or \ac{DNN} classifiers.\cite{singh2023contrastive,huang2021moltrans} 
To improve the benchmarking process, we propose to add \acp{GBM} as a baseline for \ac{DTI} benchmarking purposes, as shown in the final ablation configuration. 
\Acp{GBM} such as XGBoost have demonstrated broad adaptability, e.g. in \ac{QSAR} modelling, offering strong predictive performance and fast training times, particularly in scenarios with limited data availability, such as \ac{DTI} prediction.

We compared the overall model performance across all datasets in \cref{fig:absolute_performance_decrease}c and \cref{tab:metrics,tab:metrics_nature,tab:p_baseline}.
Here, the performance of XGBoost trained on \acp{ECFP} and \ac{PLM} embeddings is highlighted as it shows competitive performance across all methods and datasets.

\paragraph{Demonstration of the capabilities of \BTDTIXXLwospace}
To use \BTDTI in real-world applications, more training data is needed to predict meaningful interactions.
For this purpose, we have built \BTDTIXXLwospace, which is trained on more than \num{3600000} curated \ac{DTI} pairs.\cite{golts2024large}
We looked at several co-crystal structures as case studies to provide insight into the the possibilities using \BTDTIXXLwospace.
In order to demonstrate the ability to generalise beyond the learnt \acp{DTI}, we evaluated our approach on structures which are not part of the training set.
Our aim is to demonstrate the applicability of the model to multiple structures and affinities, as in the study performed by \citeauthor{dienemann2023chemical}.
The importance of this work is further emphasised by its relevance to the malaria-causing parasite \textit{Plasmodium falciparum}.\cite{dienemann2023chemical}

We first analysed the co-crystal structures \textit{Plasmodium falciparum} lipoate protein ligase 1 LipL1 (\href{https://doi.org/10.2210/pdb5T8U/pdb}{5T8U}) and \textit{Listeria monocytogenes} lplA1 (\href{https://doi.org/10.2210/pdb8CRI/pdb}{8CRI}), which share a low sequence identity (\SI{28.7}{\percent}) despite their structural similarity. 
Our objective is to evaluate the model's ability to generalise, particularly when only \ac{1D} input is provided. 
This evaluation focuses on the model's performance in capturing both biological function and structural attributes under these conditions.
Secondly, we examined the predictive shifts induced by ligand methylation and explored the interaction dynamics of a novel enzyme inhibitor C3 (\href{https://doi.org/10.2210/pdb8CRL/pdb}{8CRL}). 
This case study is further enriched with \ac{ITC} data,\cite{dienemann2023chemical} offering insights into the ligand's affinity towards the target proteins.

\begin{figure}[!hp]
    \centering
    \includegraphics[width=\textwidth]{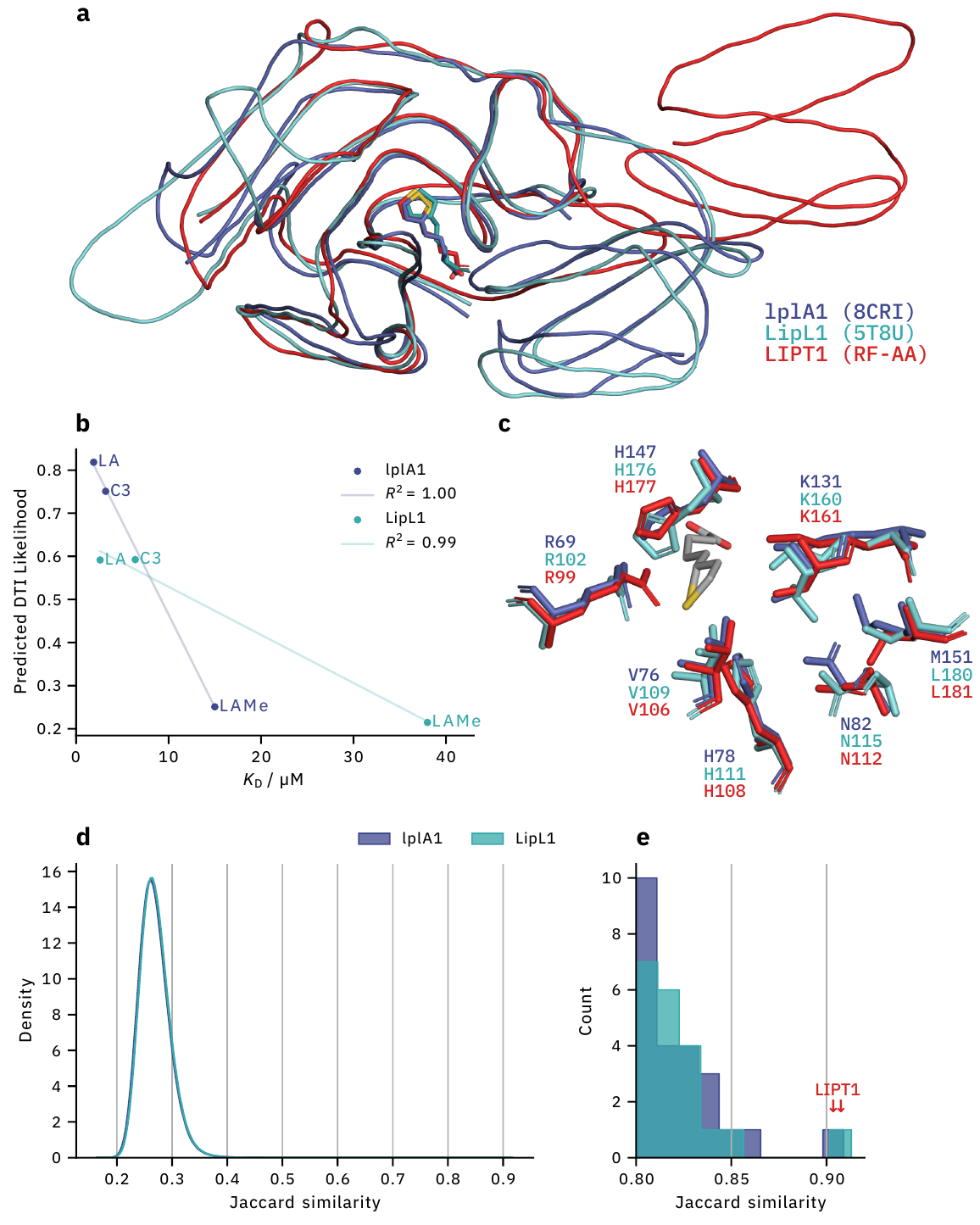}
    \caption{\textbf{Structure-based explanation of \BTDTIXXL predictions.}
    \textbf{a)} Co-crystal structures of lplA1 and LipL1 with \ac{LA} as ligand are shown in superposition, together with the most influential training sample. 
    \textbf{b)} The squared Pearson $R$\cite{pearson1895note} correlation of \BTDTIXXL and \acs{ITC} measurements is presented.\cite{dienemann2023chemical}
    \textbf{c)} The protein residue--ligand interactions at the active site are compared.
    \textbf{d)} We identified the most influential training samples for \acs{LA} predictions. The distribution of Jaccard similarity for all training samples is shown. We applied kernel density estimation to the histogram to improve visibility, due to the large training set size. 
    \textbf{e)} The most influential training samples are highlighted ($\downarrow$).
    }
    \label{fig:la_main}
\end{figure}

Our results indicate, that \BTDTIXXL is able to accurately predict the correlation between the experimentally determined affinity measured via \ac{ITC} and the likelihood of the \ac{DTI} (\cref{fig:la_main}b).
These capabilities provide useful insight in the drug discovery process, as researchers are able to prioritise chemical scaffolds.
\BTDTIXXL is able to catch small changes to the ligands structure and accurately predict the shift in interaction likelihood. 
This is illustrated by the methylation of \ac{LA}, where our method predicts a significant decrease in interaction likelihood, consistent with the decrease in affinity measured by \ac{ITC}.

We looked at \ac{SHAP} values to examine the influence of each input modality on the model (\cref{fig:lplA1_shap}).
Regardless of the ligand molecule chosen, each modality proved equally important for prediction.
This finding highlights the functionality and predictive power of \BTDTIwospace's architecture.

\paragraph{Explaining \BTDTI by investigating sample importance}
We analysed the importance of individual samples within the training set to understand how \BTDTI classifies \acp{DTI}. 
In \cref{fig:la_main}d,e, we identified the most influential training pairs by examining those with the highest Jaccard similarity, calculated from the leaf indices of the \ac{GBM} in \BTDTIXXLwospace.
The most influential training sample is the \textit{Homo sapiens} lipoyl amidotransferase LIPT1 for both lplA1 and LipL1, with \ac{LA} as the common ligand (\cref{fig:la_main}a,e). 
LIPT1 and lplA1 ($J = 0.909$) share a sequence identity of \SI{31.8}{\percent}, while LIPT1 and LipL1 ($J = 0.913$) only share \SI{29.7}{\percent} (\cref{fig:seq_align}).

To investigate the biochemical implications of the training sample to the model's prediction, we performed a structural study.
We leveraged the availability of crystallographic data to perform in-depth structural analyses on lplA1 (\href{https://doi.org/10.2210/pdb8CRI/pdb}{8CRI}) and LipL1 (\href{https://doi.org/10.2210/pdb5T8U/pdb}{5T8U}). 
A superposition of lplA1 with LIPT1 revealed a \ac{RMSD} of \SI{2.07}{\angstrom}, while LipL1 exhibited an \ac{RMSD} of \SI{1.72}{\angstrom}. 
These \ac{RMSD} values reflect a significant structural congruence among these enzymes, notwithstanding their low sequence identity. 
Despite this structural similarity, it is noteworthy that human LIPT1 does not catalyse the same reaction as lplA1 and LipL1.\cite{cao2018protein}

Furthermore, we looked at the active site of LipL1, where all residues are conserved relative to LIPT1 (\cref{fig:la_main}c).
In lplA1, one notable substitution can be observed.
L181 in LIPT1 is replaced by M151, possibly explaining the higher Jaccard similarity of LipL1 over lplA1.
This conservation pattern underscores a highly conserved binding pocket across species, as confirmed by sequence alignment data.
These results highlight the awareness of \BTDTIXXL to ligand-binding residues and help to understand how the prediction of the model is achieved.

In summary, \BTDTIXXL effectively learns \acp{DTI} by leveraging catalytically active and stabilising residues, demonstrating the model's ability to generalise from \ac{1D} input data. 
This capability makes \BTDTIXXL well-suited for applications in drug discovery.

\section{Conclusions}

Our proposed method, \BTDTIwospace, integrates sequence information with the Barlow Twins \ac{SSL} architecture and \ac{GBM} models, representing a powerful fusion of \ac{ML} and \ac{DL} techniques.

Our approach demonstrates state-of-the-art \ac{DTI} prediction capabilities, validated across multiple benchmarks and data splits. 
Notably, our method outperforms existing literature benchmarks in ten out of twelve datasets evaluated.

To elucidate the efficacy of \BTDTIwospace, we conducted an ablation study to investigate the contribution of its core components and their impact on performance. 
In addition, we re-evaluated the choice of baselines in numerous publications and advocate the inclusion of \ac{GBM} baselines.
Furthermore, we explored the classification mechanism of \BTDTI for \acp{DTI} by performing a structure-based analysis of the most influential training samples.
This was done by adapting a previously developed influence method to gain deeper insight into training sample importance.

Given the model's exceptional performance, we are confident that \BTDTI can significantly accelerate the drug discovery process and offer significant time and cost savings through the use of virtual screening campaigns.
To make \BTDTI accessible to the scientific community, we provide an easy to use and free web interface at \url{https://www.bio.nat.tum.de/oc2/barlowdti}.

\section{Methods}

\subsection{Datasets}

To evaluate the performance of \BTDTIwospace, three established benchmarks are used.
They all provide fixed splits for training, evaluation and testing.
In some publications the training and evaluation is merged to improve predictive performance.
To endure comparability, this was not done in this work.
All metrics listed from other publications are also listed where only the training set is used.

In addition, \citeauthor{kang2022finetuning} first proposed splits for large \ac{DTI} datasets, BioSNAP,\cite{biosnapnets} BindingDB\cite{liu2007bindingdb} and DAVIS.\cite{davis2011comprehensive,kang2022finetuning}

The addition of a variety of splits with an additional benchmark Human\cite{huang2021moltrans} are proposed by \citeauthor{koh2024physicochemical}, we evaluate these separately.\cite{koh2024physicochemical}

For all datasets, to reduce bias and improve model performance, the \ac{SMILES} are cleaned using the Python ChEMBL curation pipeline.\cite{bento2020open}
All duplicate and erroneous molecule and protein information that could not be parsed is removed.
Training is performed on the predefined training splits.

\subsection{Representations}

\paragraph{Molecular information}
The \ac{SMILES} are converted into \acp{ECFP} using RDKit.\cite{landrum2020rdkit}
We used them with \SI{1024}{\bit} and a radius of 2.

\paragraph{Amino acid sequence information}
The amino acid sequences are converted into vectors, by using the \ac{PLM} ProstT5.\cite{heinzinger2024bilingual}

\subsection{Barlow Twins model configuration}

The proposed method is based on the Barlow Twins\cite{zbontar2021barlow} network architecture, which employs one encoder for each modality and a unified projector.
The encoders and projector are \ac{MLP} based.
The loss function is adapted from the original Barlow Twins publication and enforces cross-correlation between the projections of the modalities.\cite{zbontar2021barlow}

The \BTDTI architecture is coded in Python using PyTorch.\cite{vanrossum1995python,paszke2019pytorch}

\paragraph{Pre-training Barlow Twins} 
Here we pre-train the Barlow Twins architecture on our joint \ac{DTI} dataset, based on BioSNAP, BindingDB, DAVIS and DrugBank,\cite{knox2024drugbank} removing duplicates and without labels to teach \acp{DTI}. 
Early stopping is implemented to avoid overfitting, which is carried out using a \SI{15}{\percent} validation split.

\subparagraph{Hyperparameter optimisation}

Manual hyperparameter optimisation is performed, shown in \cref{tab:bt_hyperparameters}.

\begin{table}[h!]
\centering
\caption{\textbf{Barlow Twins hyperparameters.} The best values are marked in bold.}
\label{tab:bt_hyperparameters}
\begin{tabular}{lS}
\toprule
\textbf{Hyperparameter} & \textbf{Value / Range} \\
\midrule
enc\_n\_neurons       & {1024, 2048, \textbf{4096}} \\
enc\_n\_layers        & {1, 2, \textbf{3}} \\
proj\_n\_neurons      & {1024, \textbf{2048}, 4096} \\
proj\_n\_layers       & {\textbf{1}, 2, 3} \\
embedding\_dim        & {\textbf{512}, 1024, 2048} \\
act\_function         & {ReLu} \\
aa\_emb\_size         & 1024 \\
loss\_weight          & {\num{1e-5}, \textbf{0.005}, 0.1} \\
batch\_size           & 4096 \\
epochs                & 250 \\
optimizer             & {AdamW} \\
learning\_rate        & {\num{1e-5}, \textbf{\num[text-series-to-math]{3e-4}}, \num{0.1}} \\
beta\_1              & 0.9 \\
beta\_2              & 0.999 \\
weight\_decay         & 5e-5 \\
step\_size            & 10 \\
gamma                & 0.1 \\
val\_split            & 0.1 \\
\bottomrule
\end{tabular}
\end{table}

\paragraph{Feature-extractor}
When performing feature-extraction, we use the pre-trained \BTDTI model.
For training and prediction, we extract the embeddings after the encoders for each modality and concatenate them.
Finally, a \ac{GBM}, XGBoost\cite{chen2016xgboost} Python implementation, is trained on the embeddings in combination with the labels for each training sets respectively.

\subparagraph{Hyperparameter optimisation}
If a benchmark provides a dedicated validation set, this was used for Optuna\cite{akiba2019optuna} hyperparameter optimisation.
The optimisation was carried out for 100 trials with the parameters shown in \cref{tab:gbm_hyperparameters}.

\begin{table}[h!]
\centering
\caption{\textbf{\Acs{GBM} hyperparameters.} Best parameters differ for each benchmarking dataset and split.}
\label{tab:gbm_hyperparameters}
\begin{tabular}{ll}
\toprule
\textbf{Hyperparameter} & \textbf{Value / Range} \\
\midrule
n\_estimators       & [100, 1000] (step=100) \\
learning\_rate      & [1e-8, 1.0] (log scale) \\
max\_depth          & [2, 12] \\
gamma               & [1e-8, 1.0] (log scale) \\
min\_child\_weight  & [1e-8, 1e2] (log scale) \\
subsample           & [0.4, 1.0] \\
reg\_lambda         & [1e-6, 10] (log scale) \\
\bottomrule
\end{tabular}
\end{table}

\subsection{\BTDTIXXLwospace}
We introduce \BTDTIXXLwospace, a model trained for use in real-world applications. 
To build \BTDTIXXLwospace, we curated and standardised the large \ac{DTI} dataset proposed by \citeauthor{golts2024large} (procedure adapted from the \enquote{Datasets} section).\cite{golts2024large}
Furthermore, we used random undersampling with a 3:1 ratio of non-interactors to interactors to improve model generalisation.
Then we added the training splits from BioSNAP, BindingDB and DAVIS, resulting in a model trained with \num{3653631} \ac{DTI} pairs (\num{2789498} non-interactors, \num{864133} interactors).

\BTDTIXXL uses the same architecture as \BTDTIwospace, using the powerful Barlow Twins network as feature-extraction method in combination with the \ac{GBM} XGBoost.\cite{zbontar2021barlow,chen2016xgboost}

\subsection{Baseline model configuration}

As a baseline, we have selected a \ac{GBM}.
Similar to our feature-extraction implementation, for all features we concatenate both \ac{ECFP} and \ac{PLM} embeddings.
Finally, a \ac{GBM}, XGBoost Python implementation, is trained on the \ac{ECFP} and \ac{PLM} embedding concatenation in combination with the labels for each training set, respectively.

\subsection{Case study}

Amino acid sequence information as well as ligand information is taken from The Protein Data Bank to perform predictions using \BTDTIwospace.\cite{berman2000protein}
Complex structures were generated using RoseTTAFold All-Atom.\cite{krishna2024generalized}

Sequence identity was determined. 
Therefore, sequences were aligned using the BLASTP\cite{altschul1990basic,altschul1997gapped} algorithm at \url{https://blast.ncbi.nlm.nih.gov}.\cite{sayers2022database}
PyMOL 2 is used for structure visualisation and \ac{RMSD} value calculation.\cite{PyMOL}

\paragraph{Explainability based on \acl{SHAP} values}
We applied the \texttt{TreeExplainer}\cite{lundberg2017unifieda,lundberg2020locala} algorithm to the \ac{GBM} of \BTDTIXXL extracted and visualised the \ac{SHAP} values.

\paragraph{Explainability based on sample importance}
To assess how the model decides to classify drug--target pairs as interacting or non-interacting, we looked at the influence of training samples, as similarly proposed by \citeauthor{brophy2022instancebased} for uncertainty estimation.\cite{brophy2022instancebased}
We used a similar concept but changed the approach to identify the most influential training data.
This is done by obtaining the leaf indices of the \ac{GBM} of all training samples.
Then we compare the leaf indices at inference time with the leaf indices of the training samples.
Finally, we find the most influential samples by computing the pairwise Jaccard similarity of the leaf index vectors,\cite{jaccard1901etude}
\[J(A, B)=\frac{|A \cap B|}{|A \cup B|}.\]
The most influential training sample is represented by the maximum Jaccard similarity.

\section{Code and Data Availability}

The easy-to-use web interface can be found at \url{https://www.bio.nat.tum.de/oc2/barlowdti}.
The code is available on GitHub \url{https://github.com/maxischuh/BarlowDTI}.
Also available on GitHub are the curated and extensive \BTDTIXXL training, as well as the benchmark data.

The system used for computational work is equipped with an AMD Ryzen Threadripper PRO 5995WX \acs{CPU} with 64/128 cores/threads and \SI{1024}{\giga\byte} \acs{RAM}.
The server is also powered by an NVIDIA RTX 4090 \acs{GPU} with \SI{24}{\giga\byte} V\acs{RAM}.

\begin{acknowledgements}
The authors thank Merck KGaA Darmstadt for their generous support with the Merck Future Insight Prize 2020. 
This project is also cofunded by the European Union (ERC, breakingBAC, 101096911).
All authors thank Prof. Michael Groll for his insight into the crystal structure data.
M.G.S. thanks Joshua Hesse and Aleksandra Daniluk for their valuable input and helpful feedback and Leonard Gareis for assistance with the website.
\end{acknowledgements}

\bibliography{ref.bib}

\begin{thebibliography}{68}
\providecommand{\natexlab}[1]{#1}
\providecommand{\url}[1]{\texttt{#1}}
\expandafter\ifx\csname urlstyle\endcsname\relax
  \providecommand{\doi}[1]{doi: #1}\else
  \providecommand{\doi}{doi: \begingroup \urlstyle{rm}\Url}\fi

\bibitem[Rang et~al.(2011)Rang, Dale, Ritter, Flower, and
  Henderson]{rang2011rang}
Humphrey~P. Rang, Maureen~M. Dale, James~M. Ritter, Rod~J. Flower, and Graeme
  Henderson.
\newblock \emph{Rang \& {{Dale}}'s {{Pharmacology}}}.
\newblock Elsevier Health Sciences, April 2011.
\newblock ISBN 978-0-7020-4504-2.

\bibitem[Strittmatter(2014)]{strittmatter2014overcoming}
Stephen~M. Strittmatter.
\newblock Overcoming {{Drug Development Bottlenecks With Repurposing}}: {{Old}}
  drugs learn new tricks.
\newblock \emph{Nature Medicine}, 20\penalty0 (6):\penalty0 590--591, June
  2014.
\newblock ISSN 1546-170X.
\newblock \doi{10.1038/nm.3595}.

\bibitem[Hughes et~al.(2011)Hughes, Rees, Kalindjian, and
  Philpott]{hughes2011principles}
Jp~Hughes, S~Rees, Sb~Kalindjian, and Kl~Philpott.
\newblock Principles of early drug discovery.
\newblock \emph{British Journal of Pharmacology}, 162\penalty0 (6):\penalty0
  1239--1249, 2011.
\newblock ISSN 1476-5381.
\newblock \doi{10.1111/j.1476-5381.2010.01127.x}.

\bibitem[Blundell et~al.(2006)Blundell, Sibanda, Montalv{\~a}o, Brewerton,
  Chelliah, Worth, Harmer, Davies, and Burke]{blundell2006structural}
Tom~L Blundell, Bancinyane~L Sibanda, Rinaldo~Wander Montalv{\~a}o, Suzanne
  Brewerton, Vijayalakshmi Chelliah, Catherine~L Worth, Nicholas~J Harmer, Owen
  Davies, and David Burke.
\newblock Structural biology and bioinformatics in drug design: Opportunities
  and challenges for target identification and lead discovery.
\newblock \emph{Philosophical Transactions of the Royal Society B: Biological
  Sciences}, 361\penalty0 (1467):\penalty0 413--423, February 2006.
\newblock \doi{10.1098/rstb.2005.1800}.

\bibitem[Tautermann(2020)]{tautermann2020current}
Christofer~S. Tautermann.
\newblock Current and {{Future Challenges}} in {{Modern Drug Discovery}}.
\newblock In Alexander Heifetz, editor, \emph{Quantum {{Mechanics}} in {{Drug
  Discovery}}}, pages 1--17. Springer US, New York, NY, 2020.
\newblock ISBN 978-1-07-160282-9.
\newblock \doi{10.1007/978-1-0716-0282-9_1}.

\bibitem[Agu et~al.(2023)Agu, Afiukwa, Orji, Ezeh, Ofoke, Ogbu, Ugwuja, and
  Aja]{agu2023molecular}
P.~C. Agu, C.~A. Afiukwa, O.~U. Orji, E.~M. Ezeh, I.~H. Ofoke, C.~O. Ogbu,
  E.~I. Ugwuja, and P.~M. Aja.
\newblock Molecular docking as a tool for the discovery of molecular targets of
  nutraceuticals in diseases management.
\newblock \emph{Scientific Reports}, 13\penalty0 (1):\penalty0 13398, August
  2023.
\newblock ISSN 2045-2322.
\newblock \doi{10.1038/s41598-023-40160-2}.

\bibitem[Bender et~al.(2021)Bender, Gahbauer, Luttens, Lyu, Webb, Stein, Fink,
  Balius, Carlsson, Irwin, and Shoichet]{bender2021practical}
Brian~J. Bender, Stefan Gahbauer, Andreas Luttens, Jiankun Lyu, Chase~M. Webb,
  Reed~M. Stein, Elissa~A. Fink, Trent~E. Balius, Jens Carlsson, John~J. Irwin,
  and Brian~K. Shoichet.
\newblock A practical guide to large-scale docking.
\newblock \emph{Nature Protocols}, 16\penalty0 (10):\penalty0 4799--4832,
  October 2021.
\newblock ISSN 1750-2799.
\newblock \doi{10.1038/s41596-021-00597-z}.

\bibitem[Hollingsworth and Dror(2018)]{hollingsworth2018molecular}
Scott~A. Hollingsworth and Ron~O. Dror.
\newblock Molecular {{Dynamics Simulation}} for {{All}}.
\newblock \emph{Neuron}, 99\penalty0 (6):\penalty0 1129--1143, September 2018.
\newblock ISSN 0896-6273.
\newblock \doi{10.1016/j.neuron.2018.08.011}.

\bibitem[Karplus and Petsko(1990)]{karplus1990molecular}
Martin Karplus and Gregory~A. Petsko.
\newblock Molecular dynamics simulations in biology.
\newblock \emph{Nature}, 347\penalty0 (6294):\penalty0 631--639, October 1990.
\newblock ISSN 1476-4687.
\newblock \doi{10.1038/347631a0}.

\bibitem[Dhakal et~al.(2022)Dhakal, McKay, Tanner, and
  Cheng]{dhakal2022artificial}
Ashwin Dhakal, Cole McKay, John~J. Tanner, and Jianlin Cheng.
\newblock Artificial intelligence in the prediction of protein--ligand
  interactions: Recent advances and future directions.
\newblock \emph{Briefings in Bioinformatics}, 23\penalty0 (1), January 2022.
\newblock \doi{10.1093/bib/bbab476}.

\bibitem[You et~al.(2022)You, Lai, Pan, Zheng, Vera, Liu, Deng, and
  Zhang]{you2022artificial}
Yujie You, Xin Lai, Yi~Pan, Huiru Zheng, Julio Vera, Suran Liu, Senyi Deng, and
  Le~Zhang.
\newblock Artificial intelligence in cancer target identification and drug
  discovery.
\newblock \emph{Signal Transduction and Targeted Therapy}, 7\penalty0
  (1):\penalty0 1--24, May 2022.
\newblock ISSN 2059-3635.
\newblock \doi{10.1038/s41392-022-00994-0}.

\bibitem[Kitchen et~al.(2004)Kitchen, Decornez, Furr, and
  Bajorath]{kitchen2004docking}
Douglas~B. Kitchen, H{\'e}l{\`e}ne Decornez, John~R. Furr, and J{\"u}rgen
  Bajorath.
\newblock Docking and scoring in virtual screening for drug discovery: Methods
  and applications.
\newblock \emph{Nature Reviews Drug Discovery}, 3\penalty0 (11):\penalty0
  935--949, November 2004.
\newblock ISSN 1474-1784.
\newblock \doi{10.1038/nrd1549}.

\bibitem[Hopkins(2009)]{hopkins2009predicting}
Andrew~L. Hopkins.
\newblock Predicting promiscuity.
\newblock \emph{Nature}, 462\penalty0 (7270):\penalty0 167--168, November 2009.
\newblock ISSN 1476-4687.
\newblock \doi{10.1038/462167a}.

\bibitem[Chen et~al.(2020)Chen, Tan, Wang, Zhong, Liu, Yang, Luo, Chen, Jiang,
  and Zheng]{chen2020transformercpi}
Lifan Chen, Xiaoqin Tan, Dingyan Wang, Feisheng Zhong, Xiaohong Liu, Tianbiao
  Yang, Xiaomin Luo, Kaixian Chen, Hualiang Jiang, and Mingyue Zheng.
\newblock {{TransformerCPI}}: Improving compound--protein interaction
  prediction by sequence-based deep learning with self-attention mechanism and
  label reversal experiments.
\newblock \emph{Bioinformatics}, 36\penalty0 (16):\penalty0 4406--4414, August
  2020.
\newblock ISSN 1367-4803.
\newblock \doi{10.1093/bioinformatics/btaa524}.

\bibitem[Jiang et~al.(2020)Jiang, Li, Zhang, Wang, Wang, Yuan, and
  Wei]{jiang2020drug}
Mingjian Jiang, Zhen Li, Shugang Zhang, Shuang Wang, Xiaofeng Wang, Qing Yuan,
  and Zhiqiang Wei.
\newblock Drug--target affinity prediction using graph neural network and
  contact maps.
\newblock \emph{RSC Advances}, 10\penalty0 (35):\penalty0 20701--20712, May
  2020.
\newblock ISSN 2046-2069.
\newblock \doi{10.1039/D0RA02297G}.

\bibitem[Koh et~al.(2024)Koh, Nguyen, Pan, May, and
  Webb]{koh2024physicochemical}
Huan~Yee Koh, Anh T.~N. Nguyen, Shirui Pan, Lauren~T. May, and Geoffrey~I.
  Webb.
\newblock Physicochemical graph neural network for learning protein--ligand
  interaction fingerprints from sequence data.
\newblock \emph{Nature Machine Intelligence}, 6\penalty0 (6):\penalty0
  673--687, June 2024.
\newblock ISSN 2522-5839.
\newblock \doi{10.1038/s42256-024-00847-1}.

\bibitem[Lee et~al.(2024)Lee, Jun, Song, and Kim]{lee2024dlmdti}
Jonghyun Lee, Dae~Won Jun, Ildae Song, and Yun Kim.
\newblock {{DLM-DTI}}: A dual language model for the prediction of drug-target
  interaction with hint-based learning.
\newblock \emph{Journal of Cheminformatics}, 16\penalty0 (1):\penalty0 1--12,
  December 2024.
\newblock ISSN 1758-2946.
\newblock \doi{10.1186/s13321-024-00808-1}.

\bibitem[Jiang et~al.(2022)Jiang, Wang, Zhang, Zhou, Zhang, and
  Li]{jiang2022sequencebased}
Mingjian Jiang, Shuang Wang, Shugang Zhang, Wei Zhou, Yuanyuan Zhang, and Zhen
  Li.
\newblock Sequence-based drug-target affinity prediction using weighted graph
  neural networks.
\newblock \emph{BMC Genomics}, 23\penalty0 (1):\penalty0 449, June 2022.
\newblock ISSN 1471-2164.
\newblock \doi{10.1186/s12864-022-08648-9}.

\bibitem[Ahdritz et~al.(2024)Ahdritz, Bouatta, Floristean, Kadyan, Xia,
  Gerecke, O'Donnell, Berenberg, Fisk, Zanichelli, Zhang, Nowaczynski, Wang,
  {Stepniewska-Dziubinska}, Zhang, Ojewole, Guney, Biderman, Watkins, Ra,
  Lorenzo, Nivon, Weitzner, Ban, Chen, Zhang, Li, Song, He, Sorger, Mostaque,
  Zhang, Bonneau, and AlQuraishi]{ahdritz2024openfolda}
Gustaf Ahdritz, Nazim Bouatta, Christina Floristean, Sachin Kadyan, Qinghui
  Xia, William Gerecke, Timothy~J. O'Donnell, Daniel Berenberg, Ian Fisk,
  Niccol{\`o} Zanichelli, Bo~Zhang, Arkadiusz Nowaczynski, Bei Wang, Marta~M.
  {Stepniewska-Dziubinska}, Shang Zhang, Adegoke Ojewole, Murat~Efe Guney,
  Stella Biderman, Andrew~M. Watkins, Stephen Ra, Pablo~Ribalta Lorenzo, Lucas
  Nivon, Brian Weitzner, Yih-En~Andrew Ban, Shiyang Chen, Minjia Zhang,
  Conglong Li, Shuaiwen~Leon Song, Yuxiong He, Peter~K. Sorger, Emad Mostaque,
  Zhao Zhang, Richard Bonneau, and Mohammed AlQuraishi.
\newblock {{OpenFold}}: Retraining {{AlphaFold2}} yields new insights into its
  learning mechanisms and capacity for generalization.
\newblock \emph{Nature Methods}, pages 1--11, May 2024.
\newblock ISSN 1548-7105.
\newblock \doi{10.1038/s41592-024-02272-z}.

\bibitem[Abramson et~al.(2024)Abramson, Adler, Dunger, Evans, Green, Pritzel,
  Ronneberger, Willmore, Ballard, Bambrick, Bodenstein, Evans, Hung, O'Neill,
  Reiman, Tunyasuvunakool, Wu, {\v Z}emgulyt{\.e}, Arvaniti, Beattie, Bertolli,
  Bridgland, Cherepanov, Congreve, {Cowen-Rivers}, Cowie, Figurnov, Fuchs,
  Gladman, Jain, Khan, Low, Perlin, Potapenko, Savy, Singh, Stecula,
  Thillaisundaram, Tong, Yakneen, Zhong, Zielinski, {\v Z}{\'i}dek, Bapst,
  Kohli, Jaderberg, Hassabis, and Jumper]{abramson2024accurate}
Josh Abramson, Jonas Adler, Jack Dunger, Richard Evans, Tim Green, Alexander
  Pritzel, Olaf Ronneberger, Lindsay Willmore, Andrew~J. Ballard, Joshua
  Bambrick, Sebastian~W. Bodenstein, David~A. Evans, Chia-Chun Hung, Michael
  O'Neill, David Reiman, Kathryn Tunyasuvunakool, Zachary Wu, Akvil{\.e} {\v
  Z}emgulyt{\.e}, Eirini Arvaniti, Charles Beattie, Ottavia Bertolli, Alex
  Bridgland, Alexey Cherepanov, Miles Congreve, Alexander~I. {Cowen-Rivers},
  Andrew Cowie, Michael Figurnov, Fabian~B. Fuchs, Hannah Gladman, Rishub Jain,
  Yousuf~A. Khan, Caroline M.~R. Low, Kuba Perlin, Anna Potapenko, Pascal Savy,
  Sukhdeep Singh, Adrian Stecula, Ashok Thillaisundaram, Catherine Tong, Sergei
  Yakneen, Ellen~D. Zhong, Michal Zielinski, Augustin {\v Z}{\'i}dek, Victor
  Bapst, Pushmeet Kohli, Max Jaderberg, Demis Hassabis, and John~M. Jumper.
\newblock Accurate structure prediction of biomolecular interactions with
  {{AlphaFold}} 3.
\newblock \emph{Nature}, 630\penalty0 (8016):\penalty0 493--500, June 2024.
\newblock ISSN 1476-4687.
\newblock \doi{10.1038/s41586-024-07487-w}.

\bibitem[Krishna et~al.(2024)Krishna, Wang, Ahern, Sturmfels, Venkatesh,
  Kalvet, Lee, {Morey-Burrows}, Anishchenko, Humphreys, McHugh, Vafeados, Li,
  Sutherland, Hitchcock, Hunter, Kang, Brackenbrough, Bera, Baek, DiMaio, and
  Baker]{krishna2024generalized}
Rohith Krishna, Jue Wang, Woody Ahern, Pascal Sturmfels, Preetham Venkatesh,
  Indrek Kalvet, Gyu~Rie Lee, Felix~S. {Morey-Burrows}, Ivan Anishchenko,
  Ian~R. Humphreys, Ryan McHugh, Dionne Vafeados, Xinting Li, George~A.
  Sutherland, Andrew Hitchcock, C.~Neil Hunter, Alex Kang, Evans Brackenbrough,
  Asim~K. Bera, Minkyung Baek, Frank DiMaio, and David Baker.
\newblock Generalized biomolecular modeling and design with {{RoseTTAFold
  All-Atom}}.
\newblock \emph{Science}, 384\penalty0 (6693):\penalty0 eadl2528, March 2024.
\newblock \doi{10.1126/science.adl2528}.

\bibitem[Trott and Olson(2010)]{trott2010autodock}
Oleg Trott and Arthur~J. Olson.
\newblock {{AutoDock Vina}}: {{Improving}} the speed and accuracy of docking
  with a new scoring function, efficient optimization, and multithreading.
\newblock \emph{Journal of Computational Chemistry}, 31\penalty0 (2):\penalty0
  455--461, 2010.
\newblock ISSN 1096-987X.
\newblock \doi{10.1002/jcc.21334}.

\bibitem[Corso et~al.(2023)Corso, St{\"a}rk, Jing, Barzilay, and
  Jaakkola]{corso2023diffdock}
Gabriele Corso, Hannes St{\"a}rk, Bowen Jing, Regina Barzilay, and Tommi
  Jaakkola.
\newblock {{DiffDock}}: {{Diffusion Steps}}, {{Twists}}, and {{Turns}} for
  {{Molecular Docking}}, February 2023.

\bibitem[He et~al.(2023)He, You, Jiang, Jiang, Xu, and Cheng]{he2023alphafold2}
Xin-heng He, Chong-zhao You, Hua-liang Jiang, Yi~Jiang, H.~Eric Xu, and
  Xi~Cheng.
\newblock {{AlphaFold2}} versus experimental structures: Evaluation on {{G}}
  protein-coupled receptors.
\newblock \emph{Acta Pharmacologica Sinica}, 44\penalty0 (1):\penalty0 1--7,
  January 2023.
\newblock ISSN 1745-7254.
\newblock \doi{10.1038/s41401-022-00938-y}.

\bibitem[Li et~al.(2021)Li, Zhou, Xu, Huang, Wang, Xiong, Huang, Dou, and
  Xiong]{li2021structureaware}
Shuangli Li, Jingbo Zhou, Tong Xu, Liang Huang, Fan Wang, Haoyi Xiong, Weili
  Huang, Dejing Dou, and Hui Xiong.
\newblock Structure-aware {{Interactive Graph Neural Networks}} for the
  {{Prediction}} of {{Protein-Ligand Binding Affinity}}.
\newblock In \emph{Proceedings of the 27th {{ACM SIGKDD Conference}} on
  {{Knowledge Discovery}} \& {{Data Mining}}}, {{KDD}} '21, pages 975--985, New
  York, NY, USA, August 2021. Association for Computing Machinery.
\newblock ISBN 978-1-4503-8332-5.
\newblock \doi{10.1145/3447548.3467311}.

\bibitem[Heinzinger et~al.(2024)Heinzinger, Weissenow, Sanchez, Henkel,
  Mirdita, Steinegger, and Rost]{heinzinger2024bilingual}
Michael Heinzinger, Konstantin Weissenow, Joaquin~Gomez Sanchez, Adrian Henkel,
  Milot Mirdita, Martin Steinegger, and Burkhard Rost.
\newblock Bilingual {{Language Model}} for {{Protein Sequence}} and
  {{Structure}}, March 2024.

\bibitem[{van Kempen} et~al.(2024){van Kempen}, Kim, Tumescheit, Mirdita, Lee,
  Gilchrist, S{\"o}ding, and Steinegger]{vankempen2024fast}
Michel {van Kempen}, Stephanie~S. Kim, Charlotte Tumescheit, Milot Mirdita,
  Jeongjae Lee, Cameron L.~M. Gilchrist, Johannes S{\"o}ding, and Martin
  Steinegger.
\newblock Fast and accurate protein structure search with {{Foldseek}}.
\newblock \emph{Nature Biotechnology}, 42\penalty0 (2):\penalty0 243--246,
  February 2024.
\newblock ISSN 1546-1696.
\newblock \doi{10.1038/s41587-023-01773-0}.

\bibitem[Chen and Guestrin(2016)]{chen2016xgboost}
Tianqi Chen and Carlos Guestrin.
\newblock {{XGBoost}}: {{A Scalable Tree Boosting System}}.
\newblock In \emph{Proceedings of the 22nd {{ACM SIGKDD International
  Conference}} on {{Knowledge Discovery}} and {{Data Mining}}}, pages 785--794,
  August 2016.
\newblock \doi{10.1145/2939672.2939785}.

\bibitem[Schuh et~al.(2024)Schuh, Boldini, and Sieber]{schuh2024synergizing}
Maximilian~G. Schuh, Davide Boldini, and Stephan~A. Sieber.
\newblock Synergizing {{Chemical Structures}} and {{Bioassay Descriptions}} for
  {{Enhanced Molecular Property Prediction}} in {{Drug Discovery}}.
\newblock \emph{Journal of Chemical Information and Modeling}, 64\penalty0
  (12):\penalty0 4640--4650, June 2024.
\newblock ISSN 1549-9596.
\newblock \doi{10.1021/acs.jcim.4c00765}.

\bibitem[Zbontar et~al.(2021)Zbontar, Jing, Misra, LeCun, and
  Deny]{zbontar2021barlow}
Jure Zbontar, Li~Jing, Ishan Misra, Yann LeCun, and St{\'e}phane Deny.
\newblock Barlow {{Twins}}: {{Self-Supervised Learning}} via {{Redundancy
  Reduction}}, June 2021.

\bibitem[Barlow et~al.(1961)]{barlow1961possible}
Horace~B Barlow et~al.
\newblock Possible principles underlying the transformation of sensory
  messages.
\newblock \emph{Sensory communication}, 1\penalty0 (01):\penalty0 217--233,
  1961.

\bibitem[Golts et~al.(2024)Golts, Ratner, Shoshan, Raboh, Polaczek,
  {Ozery-Flato}, Shats, Hazan, Ravid, and Hexter]{golts2024large}
Alex Golts, Vadim Ratner, Yoel Shoshan, Moshe Raboh, Sagi Polaczek, Michal
  {Ozery-Flato}, Daniel Shats, Liam Hazan, Sivan Ravid, and Efrat Hexter.
\newblock A large dataset curation and benchmark for drug target interaction,
  January 2024.

\bibitem[Brophy and Lowd(2022)]{brophy2022instancebased}
Jonathan Brophy and Daniel Lowd.
\newblock Instance-{{Based Uncertainty Estimation}} for {{Gradient-Boosted
  Regression Trees}}, October 2022.

\bibitem[Zitnik et~al.(2018)Zitnik, Sosi\v{c}, Maheshwari, and
  Leskovec]{biosnapnets}
Marinka Zitnik, Rok Sosi\v{c}, Sagar Maheshwari, and Jure Leskovec.
\newblock {BioSNAP Datasets}: {Stanford} biomedical network dataset collection.
\newblock \url{http://snap.stanford.edu/biodata}, August 2018.

\bibitem[Liu et~al.(2007)Liu, Lin, Wen, Jorissen, and Gilson]{liu2007bindingdb}
Tiqing Liu, Yuhmei Lin, Xin Wen, Robert~N. Jorissen, and Michael~K. Gilson.
\newblock {{BindingDB}}: A web-accessible database of experimentally determined
  protein--ligand binding affinities.
\newblock \emph{Nucleic Acids Research}, 35\penalty0 (suppl\_1):\penalty0
  D198--D201, January 2007.
\newblock ISSN 0305-1048.
\newblock \doi{10.1093/nar/gkl999}.

\bibitem[Davis et~al.(2011)Davis, Hunt, Herrgard, Ciceri, Wodicka, Pallares,
  Hocker, Treiber, and Zarrinkar]{davis2011comprehensive}
Mindy~I. Davis, Jeremy~P. Hunt, Sanna Herrgard, Pietro Ciceri, Lisa~M. Wodicka,
  Gabriel Pallares, Michael Hocker, Daniel~K. Treiber, and Patrick~P.
  Zarrinkar.
\newblock Comprehensive analysis of kinase inhibitor selectivity.
\newblock \emph{Nature Biotechnology}, 29\penalty0 (11):\penalty0 1046--1051,
  November 2011.
\newblock ISSN 1546-1696.
\newblock \doi{10.1038/nbt.1990}.

\bibitem[Knox et~al.(2024)Knox, Wilson, Klinger, Franklin, Oler, Wilson, Pon,
  Cox, Chin, Strawbridge, {Garcia-Patino}, Kruger, Sivakumaran, Sanford, Doshi,
  Khetarpal, Fatokun, Doucet, Zubkowski, Rayat, Jackson, Harford, Anjum, Zakir,
  Wang, Tian, Lee, Liigand, Peters, Wang, Nguyen, So, Sharp, {da~Silva},
  Gabriel, Scantlebury, Jasinski, Ackerman, Jewison, Sajed, Gautam, and
  Wishart]{knox2024drugbank}
Craig Knox, Mike Wilson, Christen~M Klinger, Mark Franklin, Eponine Oler, Alex
  Wilson, Allison Pon, Jordan Cox, Na~Eun~(Lucy) Chin, Seth~A Strawbridge,
  Marysol {Garcia-Patino}, Ray Kruger, Aadhavya Sivakumaran, Selena Sanford,
  Rahil Doshi, Nitya Khetarpal, Omolola Fatokun, Daphnee Doucet, Ashley
  Zubkowski, Dorsa~Yahya Rayat, Hayley Jackson, Karxena Harford, Afia Anjum,
  Mahi Zakir, Fei Wang, Siyang Tian, Brian Lee, Jaanus Liigand, Harrison
  Peters, Ruo Qi~(Rachel) Wang, Tue Nguyen, Denise So, Matthew Sharp, Rodolfo
  {da~Silva}, Cyrella Gabriel, Joshua Scantlebury, Marissa Jasinski, David
  Ackerman, Timothy Jewison, Tanvir Sajed, Vasuk Gautam, and David~S Wishart.
\newblock {{DrugBank}} 6.0: The {{DrugBank Knowledgebase}} for 2024.
\newblock \emph{Nucleic Acids Research}, 52\penalty0 (D1):\penalty0
  D1265--D1275, January 2024.
\newblock ISSN 0305-1048.
\newblock \doi{10.1093/nar/gkad976}.

\bibitem[Kim et~al.(2023)Kim, Chen, Cheng, Gindulyte, He, He, Li, Shoemaker,
  Thiessen, Yu, Zaslavsky, Zhang, and Bolton]{kim2023pubchem}
Sunghwan Kim, Jie Chen, Tiejun Cheng, Asta Gindulyte, Jia He, Siqian He,
  Qingliang Li, Benjamin~A. Shoemaker, Paul~A. Thiessen, Bo~Yu, Leonid
  Zaslavsky, Jian Zhang, and Evan~E. Bolton.
\newblock {{PubChem}} 2023 update.
\newblock \emph{Nucleic Acids Research}, 51\penalty0 (D1):\penalty0
  D1373--D1380, January 2023.
\newblock ISSN 1362-4962.
\newblock \doi{10.1093/nar/gkac956}.

\bibitem[Mendez et~al.(2019)Mendez, Gaulton, Bento, Chambers, De~Veij,
  F{\'e}lix, Magari{\~n}os, Mosquera, Mutowo, Nowotka,
  {Gordillo-Mara{\~n}{\'o}n}, Hunter, Junco, Mugumbate, {Rodriguez-Lopez},
  Atkinson, Bosc, Radoux, {Segura-Cabrera}, Hersey, and
  Leach]{mendez2019chembl}
David Mendez, Anna Gaulton, A~Patr{\'i}cia Bento, Jon Chambers, Marleen
  De~Veij, Eloy F{\'e}lix, Mar{\'i}a~Paula Magari{\~n}os, Juan~F Mosquera,
  Prudence Mutowo, Michal Nowotka, Mar{\'i}a {Gordillo-Mara{\~n}{\'o}n}, Fiona
  Hunter, Laura Junco, Grace Mugumbate, Milagros {Rodriguez-Lopez}, Francis
  Atkinson, Nicolas Bosc, Chris~J Radoux, Aldo {Segura-Cabrera}, Anne Hersey,
  and Andrew~R Leach.
\newblock {{ChEMBL}}: Towards direct deposition of bioassay data.
\newblock \emph{Nucleic acids research}, 47\penalty0 (D1):\penalty0 D930--D940,
  January 2019.
\newblock ISSN 1362-4962.
\newblock \doi{10.1093/nar/gky1075}.

\bibitem[Kang et~al.(2022)Kang, Goo, Lee, Chae, Yun, and
  Jung]{kang2022finetuning}
Hyeunseok Kang, Sungwoo Goo, Hyunjung Lee, Jung-woo Chae, Hwi-yeol Yun, and
  Sangkeun Jung.
\newblock Fine-tuning of {{BERT Model}} to {{Accurately Predict
  Drug}}--{{Target Interactions}}.
\newblock \emph{Pharmaceutics}, 14\penalty0 (8):\penalty0 1710, August 2022.
\newblock ISSN 1999-4923.
\newblock \doi{10.3390/pharmaceutics14081710}.

\bibitem[Huang et~al.(2021)Huang, Xiao, Glass, and Sun]{huang2021moltrans}
Kexin Huang, Cao Xiao, Lucas~M Glass, and Jimeng Sun.
\newblock {{MolTrans}}: {{Molecular Interaction Transformer}} for drug--target
  interaction prediction.
\newblock \emph{Bioinformatics}, 37\penalty0 (6):\penalty0 830--836, March
  2021.
\newblock ISSN 1367-4803.
\newblock \doi{10.1093/bioinformatics/btaa880}.

\bibitem[Singh et~al.(2023)Singh, Sledzieski, Bryson, Cowen, and
  Berger]{singh2023contrastive}
Rohit Singh, Samuel Sledzieski, Bryan Bryson, Lenore Cowen, and Bonnie Berger.
\newblock Contrastive learning in protein language space predicts interactions
  between drugs and protein targets.
\newblock \emph{Proceedings of the National Academy of Sciences}, 120\penalty0
  (24):\penalty0 e2220778120, June 2023.
\newblock \doi{10.1073/pnas.2220778120}.

\bibitem[Bai et~al.(2023)Bai, Miljkovi{\'c}, John, and
  Lu]{bai2023interpretable}
Peizhen Bai, Filip Miljkovi{\'c}, Bino John, and Haiping Lu.
\newblock Interpretable bilinear attention network with domain adaptation
  improves drug--target prediction.
\newblock \emph{Nature Machine Intelligence}, 5\penalty0 (2):\penalty0
  126--136, February 2023.
\newblock ISSN 2522-5839.
\newblock \doi{10.1038/s42256-022-00605-1}.

\bibitem[Wang et~al.(2022)Wang, Zheng, Jiang, Li, Liu, Wen, Patronov, Qian,
  Chen, and Yang]{wang2022structureaware}
Penglei Wang, Shuangjia Zheng, Yize Jiang, Chengtao Li, Junhong Liu, Chang Wen,
  Atanas Patronov, Dahong Qian, Hongming Chen, and Yuedong Yang.
\newblock Structure-{{Aware Multimodal Deep Learning}} for {{Drug}}--{{Protein
  Interaction Prediction}}.
\newblock \emph{Journal of Chemical Information and Modeling}, 62\penalty0
  (5):\penalty0 1308--1317, March 2022.
\newblock ISSN 1549-9596.
\newblock \doi{10.1021/acs.jcim.2c00060}.

\bibitem[Wu et~al.(2021)Wu, Zhu, Kang, Leung, Lei, Shen, Jiang, Wang, Cao, and
  Hou]{wu2021we}
Zhenxing Wu, Minfeng Zhu, Yu~Kang, Elaine Lai-Han Leung, Tailong Lei, Chao
  Shen, Dejun Jiang, Zhe Wang, Dongsheng Cao, and Tingjun Hou.
\newblock Do we need different machine learning algorithms for {{QSAR}}
  modeling? {{A}} comprehensive assessment of 16 machine learning algorithms on
  14 {{QSAR}} data sets.
\newblock \emph{Briefings in Bioinformatics}, 22\penalty0 (4):\penalty0
  bbaa321, July 2021.
\newblock ISSN 1477-4054.
\newblock \doi{10.1093/bib/bbaa321}.

\bibitem[Sheridan et~al.(2016)Sheridan, Wang, Liaw, Ma, and
  Gifford]{sheridan2016extreme}
Robert~P. Sheridan, Wei~Min Wang, Andy Liaw, Junshui Ma, and Eric~M. Gifford.
\newblock Extreme {{Gradient Boosting}} as a {{Method}} for {{Quantitative
  Structure}}--{{Activity Relationships}}.
\newblock \emph{Journal of Chemical Information and Modeling}, 56\penalty0
  (12):\penalty0 2353--2360, December 2016.
\newblock ISSN 1549-9596.
\newblock \doi{10.1021/acs.jcim.6b00591}.

\bibitem[Asselman et~al.(2023)Asselman, Khaldi, and
  Aammou]{asselman2023enhancing}
Amal Asselman, Mohamed Khaldi, and Souhaib Aammou.
\newblock Enhancing the prediction of student performance based on the machine
  learning {{XGBoost}} algorithm.
\newblock \emph{Interactive Learning Environments}, 31\penalty0 (6):\penalty0
  3360--3379, August 2023.
\newblock ISSN 1049-4820.
\newblock \doi{10.1080/10494820.2021.1928235}.

\bibitem[WELCH(1947)]{welch1947generalization}
B.~L. WELCH.
\newblock {{THE GENERALIZATION OF}} `{{STUDENT}}'{{S}}' {{PROBLEM WHEN SEVERAL
  DIFFERENT POPULATION VARLANCES ARE INVOLVED}}.
\newblock \emph{Biometrika}, 34\penalty0 (1-2):\penalty0 28--35, January 1947.
\newblock ISSN 0006-3444.
\newblock \doi{10.1093/biomet/34.1-2.28}.

\bibitem[Virtanen et~al.(2020)Virtanen, Gommers, Oliphant, Haberland, Reddy,
  Cournapeau, Burovski, Peterson, Weckesser, Bright, {van der Walt}, Brett,
  Wilson, Millman, Mayorov, Nelson, Jones, Kern, Larson, Carey, Polat, Feng,
  Moore, VanderPlas, Laxalde, Perktold, Cimrman, Henriksen, Quintero, Harris,
  Archibald, Ribeiro, Pedregosa, and {van Mulbregt}]{virtanen2020scipy}
Pauli Virtanen, Ralf Gommers, Travis~E. Oliphant, Matt Haberland, Tyler Reddy,
  David Cournapeau, Evgeni Burovski, Pearu Peterson, Warren Weckesser, Jonathan
  Bright, St{\'e}fan~J. {van der Walt}, Matthew Brett, Joshua Wilson, K.~Jarrod
  Millman, Nikolay Mayorov, Andrew R.~J. Nelson, Eric Jones, Robert Kern, Eric
  Larson, C.~J. Carey, {\.I}lhan Polat, Yu~Feng, Eric~W. Moore, Jake
  VanderPlas, Denis Laxalde, Josef Perktold, Robert Cimrman, Ian Henriksen,
  E.~A. Quintero, Charles~R. Harris, Anne~M. Archibald, Ant{\^o}nio~H. Ribeiro,
  Fabian Pedregosa, and Paul {van Mulbregt}.
\newblock {{SciPy}} 1.0: Fundamental algorithms for scientific computing in
  {{Python}}.
\newblock \emph{Nature Methods}, 17\penalty0 (3):\penalty0 261--272, March
  2020.
\newblock ISSN 1548-7105.
\newblock \doi{10.1038/s41592-019-0686-2}.

\bibitem[Benjamini and Hochberg(1995)]{benjamini1995controlling}
Yoav Benjamini and Yosef Hochberg.
\newblock Controlling the {{False Discovery Rate}}: {{A Practical}} and
  {{Powerful Approach}} to {{Multiple Testing}}.
\newblock \emph{Journal of the Royal Statistical Society. Series B
  (Methodological)}, 57\penalty0 (1):\penalty0 289--300, 1995.
\newblock ISSN 0035-9246.

\bibitem[Elnaggar et~al.(2022)Elnaggar, Heinzinger, Dallago, Rehawi, Wang,
  Jones, Gibbs, Feher, Angerer, Steinegger, Bhowmik, and
  Rost]{elnaggar2022prottrans}
Ahmed Elnaggar, Michael Heinzinger, Christian Dallago, Ghalia Rehawi, Yu~Wang,
  Llion Jones, Tom Gibbs, Tamas Feher, Christoph Angerer, Martin Steinegger,
  Debsindhu Bhowmik, and Burkhard Rost.
\newblock {{ProtTrans}}: {{Toward Understanding}} the {{Language}} of {{Life
  Through Self-Supervised Learning}}.
\newblock \emph{IEEE Transactions on Pattern Analysis and Machine
  Intelligence}, 44\penalty0 (10):\penalty0 7112--7127, October 2022.
\newblock ISSN 1939-3539.
\newblock \doi{10.1109/TPAMI.2021.3095381}.

\bibitem[Dienemann et~al.(2023)Dienemann, Chen, Hitzenberger, Sievert, Hacker,
  Prigge, Zacharias, Groll, and Sieber]{dienemann2023chemical}
Jan-Niklas Dienemann, Shu-Yu Chen, Manuel Hitzenberger, Montana~L. Sievert,
  Stephan~M. Hacker, Sean~T. Prigge, Martin Zacharias, Michael Groll, and
  Stephan~A. Sieber.
\newblock A {{Chemical Proteomic Strategy Reveals Inhibitors}} of {{Lipoate
  Salvage}} in {{Bacteria}} and {{Parasites}}.
\newblock \emph{Angewandte Chemie International Edition}, 62\penalty0
  (31):\penalty0 e202304533, 2023.
\newblock ISSN 1521-3773.
\newblock \doi{10.1002/anie.202304533}.

\bibitem[Pearson(1895)]{pearson1895note}
Karl Pearson.
\newblock Note on {{Regression}} and {{Inheritance}} in the {{Case}} of {{Two
  Parents}}.
\newblock \emph{Proceedings of the Royal Society of London Series I},
  58:\penalty0 240--242, January 1895.

\bibitem[Cao et~al.(2018)Cao, Zhu, Song, Hu, and Cronan]{cao2018protein}
Xinyun Cao, Lei Zhu, Xuejiao Song, Zhe Hu, and John~E Cronan.
\newblock Protein moonlighting elucidates the essential human pathway
  catalyzing lipoic acid assembly on its cognate enzymes.
\newblock \emph{Proceedings of the National Academy of Sciences}, 115\penalty0
  (30):\penalty0 E7063--E7072, 2018.

\bibitem[Bento et~al.(2020)Bento, Hersey, F{\'e}lix, Landrum, Gaulton,
  Atkinson, Bellis, De~Veij, and Leach]{bento2020open}
A.~Patr{\'i}cia Bento, Anne Hersey, Eloy F{\'e}lix, Greg Landrum, Anna Gaulton,
  Francis Atkinson, Louisa~J. Bellis, Marleen De~Veij, and Andrew~R. Leach.
\newblock An open source chemical structure curation pipeline using {{RDKit}}.
\newblock \emph{Journal of Cheminformatics}, 12\penalty0 (1):\penalty0 51,
  September 2020.
\newblock ISSN 1758-2946.
\newblock \doi{10.1186/s13321-020-00456-1}.

\bibitem[Landrum et~al.(2020)Landrum, Tosco, Kelley, {sriniker}, {gedeck},
  NadineSchneider, Vianello, Ric, Dalke, Cole, AlexanderSavelyev, Swain, Turk,
  N, Vaucher, Kawashima, W{\'o}jcikowski, Probst, {godin}, Cosgrove, Pahl, JP,
  Berenger, {strets123}, JLVarjo, O'Boyle, Fuller, Jensen, Sforna, and
  DoliathGavid]{landrum2020rdkit}
Greg Landrum, Paolo Tosco, Brian Kelley, {sriniker}, {gedeck}, NadineSchneider,
  Riccardo Vianello, Ric, Andrew Dalke, Brian Cole, AlexanderSavelyev, Matt
  Swain, Samo Turk, Dan N, Alain Vaucher, Eisuke Kawashima, Maciej
  W{\'o}jcikowski, Daniel Probst, guillaume {godin}, David Cosgrove, Axel Pahl,
  JP, Francois Berenger, {strets123}, JLVarjo, Noel O'Boyle, Patrick Fuller,
  Jan~Holst Jensen, Gianluca Sforna, and DoliathGavid.
\newblock Rdkit/rdkit: 2020\_03\_1 ({{Q1}} 2020) {{Release}}.
\newblock Zenodo, March 2020.

\bibitem[{van Rossum}(1995)]{vanrossum1995python}
Guido {van Rossum}.
\newblock Python tutorial.
\newblock \penalty0 (R 9526), January 1995.

\bibitem[Paszke et~al.(2019)Paszke, Gross, Massa, Lerer, Bradbury, Chanan,
  Killeen, Lin, Gimelshein, Antiga, Desmaison, K{\"o}pf, Yang, DeVito, Raison,
  Tejani, Chilamkurthy, Steiner, Fang, Bai, and Chintala]{paszke2019pytorch}
Adam Paszke, Sam Gross, Francisco Massa, Adam Lerer, James Bradbury, Gregory
  Chanan, Trevor Killeen, Zeming Lin, Natalia Gimelshein, Luca Antiga, Alban
  Desmaison, Andreas K{\"o}pf, Edward Yang, Zach DeVito, Martin Raison, Alykhan
  Tejani, Sasank Chilamkurthy, Benoit Steiner, Lu~Fang, Junjie Bai, and Soumith
  Chintala.
\newblock {{PyTorch}}: {{An Imperative Style}}, {{High-Performance Deep
  Learning Library}}, December 2019.

\bibitem[Akiba et~al.(2019)Akiba, Sano, Yanase, Ohta, and
  Koyama]{akiba2019optuna}
Takuya Akiba, Shotaro Sano, Toshihiko Yanase, Takeru Ohta, and Masanori Koyama.
\newblock Optuna: {{A Next-generation Hyperparameter Optimization Framework}}.
\newblock In \emph{Proceedings of the 25th {{ACM SIGKDD International
  Conference}} on {{Knowledge Discovery}} \& {{Data Mining}}}, {{KDD}} '19,
  pages 2623--2631, New York, NY, USA, July 2019. Association for Computing
  Machinery.
\newblock ISBN 978-1-4503-6201-6.
\newblock \doi{10.1145/3292500.3330701}.

\bibitem[Berman et~al.(2000)Berman, Westbrook, Feng, Gilliland, Bhat, Weissig,
  Shindyalov, and Bourne]{berman2000protein}
Helen~M. Berman, John Westbrook, Zukang Feng, Gary Gilliland, T.~N. Bhat, Helge
  Weissig, Ilya~N. Shindyalov, and Philip~E. Bourne.
\newblock The {{Protein Data Bank}}.
\newblock \emph{Nucleic Acids Research}, 28\penalty0 (1):\penalty0 235--242,
  January 2000.
\newblock ISSN 0305-1048.
\newblock \doi{10.1093/nar/28.1.235}.

\bibitem[Altschul et~al.(1990)Altschul, Gish, Miller, Myers, and
  Lipman]{altschul1990basic}
Stephen~F. Altschul, Warren Gish, Webb Miller, Eugene~W. Myers, and David~J.
  Lipman.
\newblock Basic local alignment search tool.
\newblock \emph{Journal of Molecular Biology}, 215\penalty0 (3):\penalty0
  403--410, October 1990.
\newblock ISSN 0022-2836.
\newblock \doi{10.1016/S0022-2836(05)80360-2}.

\bibitem[Altschul et~al.(1997)Altschul, Madden, Sch{\"a}ffer, Zhang, Zhang,
  Miller, and Lipman]{altschul1997gapped}
Stephen~F. Altschul, Thomas~L. Madden, Alejandro~A. Sch{\"a}ffer, Jinghui
  Zhang, Zheng Zhang, Webb Miller, and David~J. Lipman.
\newblock Gapped {{BLAST}} and {{PSI-BLAST}}: A new generation of protein
  database search programs.
\newblock \emph{Nucleic Acids Research}, 25\penalty0 (17):\penalty0 3389--3402,
  September 1997.
\newblock ISSN 0305-1048.
\newblock \doi{10.1093/nar/25.17.3389}.

\bibitem[Sayers et~al.(2022)Sayers, Bolton, Brister, Canese, Chan, Comeau,
  Connor, Funk, Kelly, Kim, Madej, {Marchler-Bauer}, Lanczycki, Lathrop, Lu,
  {Thibaud-Nissen}, Murphy, Phan, Skripchenko, Tse, Wang, Williams, Trawick,
  Pruitt, and Sherry]{sayers2022database}
Eric~W Sayers, Evan~E Bolton, J~Rodney Brister, Kathi Canese, Jessica Chan,
  Donald~C Comeau, Ryan Connor, Kathryn Funk, Chris Kelly, Sunghwan Kim, Tom
  Madej, Aron {Marchler-Bauer}, Christopher Lanczycki, Stacy Lathrop, Zhiyong
  Lu, Francoise {Thibaud-Nissen}, Terence Murphy, Lon Phan, Yuri Skripchenko,
  Tony Tse, Jiyao Wang, Rebecca Williams, Barton~W Trawick, Kim~D Pruitt, and
  Stephen~T Sherry.
\newblock Database resources of the national center for biotechnology
  information.
\newblock \emph{Nucleic Acids Research}, 50\penalty0 (D1):\penalty0 D20--D26,
  January 2022.
\newblock ISSN 0305-1048.
\newblock \doi{10.1093/nar/gkab1112}.

\bibitem[{Schr\"odinger, LLC}(2015)]{PyMOL}
{Schr\"odinger, LLC}.
\newblock The {PyMOL} molecular graphics system, version~1.8.
\newblock November 2015.

\bibitem[Lundberg and Lee(2017)]{lundberg2017unifieda}
Scott~M Lundberg and Su-In Lee.
\newblock A {{Unified Approach}} to {{Interpreting Model Predictions}}.
\newblock In \emph{Advances in {{Neural Information Processing Systems}}},
  volume~30. Curran Associates, Inc., 2017.

\bibitem[Lundberg et~al.(2020)Lundberg, Erion, Chen, DeGrave, Prutkin, Nair,
  Katz, Himmelfarb, Bansal, and Lee]{lundberg2020locala}
Scott~M. Lundberg, Gabriel Erion, Hugh Chen, Alex DeGrave, Jordan~M. Prutkin,
  Bala Nair, Ronit Katz, Jonathan Himmelfarb, Nisha Bansal, and Su-In Lee.
\newblock From local explanations to global understanding with explainable
  {{AI}} for trees.
\newblock \emph{Nature Machine Intelligence}, 2\penalty0 (1):\penalty0 56--67,
  January 2020.
\newblock ISSN 2522-5839.
\newblock \doi{10.1038/s42256-019-0138-9}.

\bibitem[Jaccard(1901)]{jaccard1901etude}
Paul Jaccard.
\newblock {\'E}tude comparative de la distribution florale dans une portion des
  {{Alpes}} et du {{Jura}}.
\newblock \emph{Bulletin de la Soci{\'e}t{\'e} Vaudoise des Sciences
  Naturelles}, 37\penalty0 (142):\penalty0 547, 1901.
\newblock ISSN 0037-9603.
\newblock \doi{10.5169/seals-266450}.

\bibitem[Wilcoxon(1945)]{wilcoxon1945individual}
Frank Wilcoxon.
\newblock Individual {{Comparisons}} by {{Ranking Methods}}.
\newblock \emph{Biometrics Bulletin}, 1\penalty0 (6):\penalty0 80--83, 1945.
\newblock ISSN 0099-4987.
\newblock \doi{10.2307/3001968}.

\end{thebibliography}

\newpage
\appendix


\section{Additional Results and Discussion}

\paragraph{Statistical testing}
We focus on \ac{PR_AUC} as our metric because it is an established performance indicator in unbalanced scenarios. 
Secondly, it shows a more pronounced separation between different methods, as most methods show very high values of \ac{ROC_AUC}. 

We apply the two-sided Welch's $t$-test,\cite{welch1947generalization,virtanen2020scipy} with Benjamini-Hochberg\cite{benjamini1995controlling} multiple test correction.
This is done for all methods for which the required performance information exists in the published literature.

In \cref{fig:absolute_performance_decrease}, our primary focus is on the overall change in performance.
We therefore make comparisons across all datasets collectively rather than individually. 
Detailed individual comparisons are provided in \cref{tab:metrics,tab:metrics_nature}.

\begin{table}[h!]
    \centering
    \caption{Statistical testing of benchmarking \BTDTI against other models using \citeauthor{kang2022finetuning} splits.\cite{kang2022finetuning} Five replicates were performed. Two-sided Welch's $t$-test,\cite{welch1947generalization,virtanen2020scipy} $\alpha = 0.001$ with Benjamini-Hochberg\cite{benjamini1995controlling} multiple test correction was applied.}
    \label{tab:p_metrics}
    \sisetup{
		table-number-alignment = left,
		table-alignment-mode = none,
	}
    \begin{tabular}{llScSc}
        \toprule
        \textbf{Dataset} & \textbf{Model} & \multicolumn{2}{c}{\textbf{\acs{ROC_AUC}}} & \multicolumn{2}{c}{\textbf{\acs{PR_AUC}}} \\
        \cmidrule(r){3-4} \cmidrule(l){5-6}
        \multicolumn{2}{c}{} & {$p_\text{corr}$ value} & Significant & {$p_\text{corr}$ value} & Significant \\
        \midrule
        \multirow{5}{*}{BioSNAP} 
                  & XGBoost      & 5.63e-09 & True        & 4.06e-09 & True        \\
                  & MolTrans\cite{huang2021moltrans} & 1.70e-07 & True        & 4.48e-06 & True        \\
                  & \citeauthor{kang2022finetuning}     & 6.49e-05 & True        & 3.37e-05 & True        \\
                  & DLM-DTI\cite{lee2024dlmdti}  & 4.26e-06 & True        & 3.65e-05 & True        \\
                  & ConPLex\cite{singh2023contrastive}  & {---}      & {---}         & 5.77e-10 & True        \\
        \midrule
        \multirow{5}{*}{BindingDB}
                  & XGBoost      & 3.45e-07 & True        & 1.89e-06 & True        \\
                  & MolTrans\cite{huang2021moltrans} & 3.45e-07 & True        & 1.89e-06 & True        \\
                  & \citeauthor{kang2022finetuning}     & 1.70e-06 & True        & 1.19e-05 & True        \\
                  & DLM-DTI\cite{lee2024dlmdti}  & 1.58e-04 & True        & 1.89e-06 & True        \\
                  & ConPLex\cite{singh2023contrastive} & {---}      & {---}         & 2.84e-05 & True        \\
        \midrule
        \multirow{5}{*}{DAVIS} 
                  & XGBoost      & 6.78e-07 & True        & 5.79e-08 & True        \\
                  & MolTrans\cite{huang2021moltrans} & 2.89e-07 & True        & 3.74e-05 & True        \\
                  & \citeauthor{kang2022finetuning}     & 6.78e-07 & True        & 1.41e-06 & True        \\
                  & DLM-DTI\cite{lee2024dlmdti}  & 6.78e-07 & True        & 3.00e-05 & True        \\
                  & ConPLex \cite{singh2023contrastive} & {---}      & {---}         & 1.82e-04 & True        \\
        \bottomrule
    \end{tabular}
\end{table}

\begin{table}[!htb]
\centering
\caption{Statistical testing of ablation benchmark with \BTDTI against other models using \citeauthor{kang2022finetuning} splits.\cite{kang2022finetuning} Five replicates each are performed. Two-sided Welch's $t$-test,\cite{welch1947generalization,virtanen2020scipy} $\alpha = 0.0001$ with Benjamini-Hochberg\cite{benjamini1995controlling} multiple test correction was applied. (o.: optimised; n.o.: non-optimised)}
\label{tab:p_ablation}
\sisetup{
    table-number-alignment = left,
    table-alignment-mode = none,
}
\begin{tabular}{llSc}
\toprule
\multicolumn{2}{c}{\textbf{Comparison}} & \multicolumn{2}{c}{\textbf{\acs{PR_AUC}}} \\
\cmidrule(r){1-2} \cmidrule(l){3-4}
Model 1 & Model 2 & {$p_\text{corr}$ value} & Significant \\
\midrule
XGBoost & XGBoost o. & 1.47e-07 & True \\
XGBoost & \BTDTI n.o. & 2.37e-23 & True \\
XGBoost & \BTDTI & 6.32e-25 & True \\
XGBoost o. & \BTDTI n.o. & 1.64e-10 & True \\
XGBoost o. & \BTDTI & 9.48e-15 & True \\
\BTDTI n.o. & \BTDTI & 1.70e-15 & True \\
\bottomrule
\end{tabular}
\end{table}

\begin{table}[!htb]
\centering
\caption{Statistical testing of benchmarking \BTDTI against other models using \citeauthor{kang2022finetuning} splits.\cite{kang2022finetuning} For XGBoost five replicates were performed. Two-sided Welch's $t$-test,\cite{welch1947generalization,virtanen2020scipy} $\alpha = 0.05$ with Benjamini-Hochberg\cite{benjamini1995controlling} multiple test correction was applied.}
\label{tab:p_baseline}
\sisetup{
    table-number-alignment = left,
    table-alignment-mode = none,
}
\begin{tabular}{llSc}
\toprule
\multicolumn{2}{c}{\textbf{Comparison}} & \multicolumn{2}{c}{\textbf{\acs{PR_AUC}}} \\
\cmidrule(r){1-2} \cmidrule(l){3-4}
Model 1 & Model 2 & {$p_\text{corr}$ value} & Significant \\
\midrule
XGBoost & DLM-DTI & 0.1452 & False \\
XGBoost & MolTrans & 0.1452 & False \\
XGBoost & \citeauthor{kang2022finetuning} & 0.1452 & False \\
DLM-DTI & MolTrans & 0.7970 & False \\
DLM-DTI & \citeauthor{kang2022finetuning} & 0.7970 & False \\
MolTrans & \citeauthor{kang2022finetuning} & 0.7970 & False \\
\bottomrule
\end{tabular}
\end{table}

\begin{figure}[!htb]
    \centering
    \includegraphics[width=\textwidth]{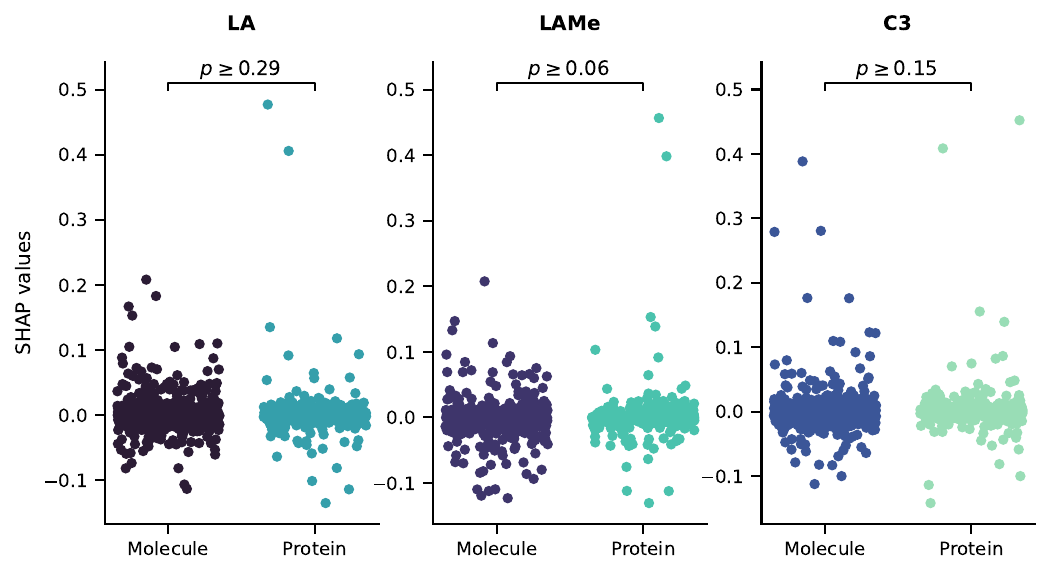}
    \caption{\acs{SHAP} values of \BTDTIXXL input modalities. No significant change in distribution could be shown, independent of the ligand molecule, case study based on the \citeauthor{dienemann2023chemical} publication.\cite{dienemann2023chemical} A two-sided Wilcoxon\cite{wilcoxon1945individual} signed-rank test was applied and respective $p$-values are presented within the figure.}
    \label{fig:lplA1_shap}
\end{figure}

\begin{figure}[!htb]
    \centering
    \includegraphics[width=.8\textwidth]{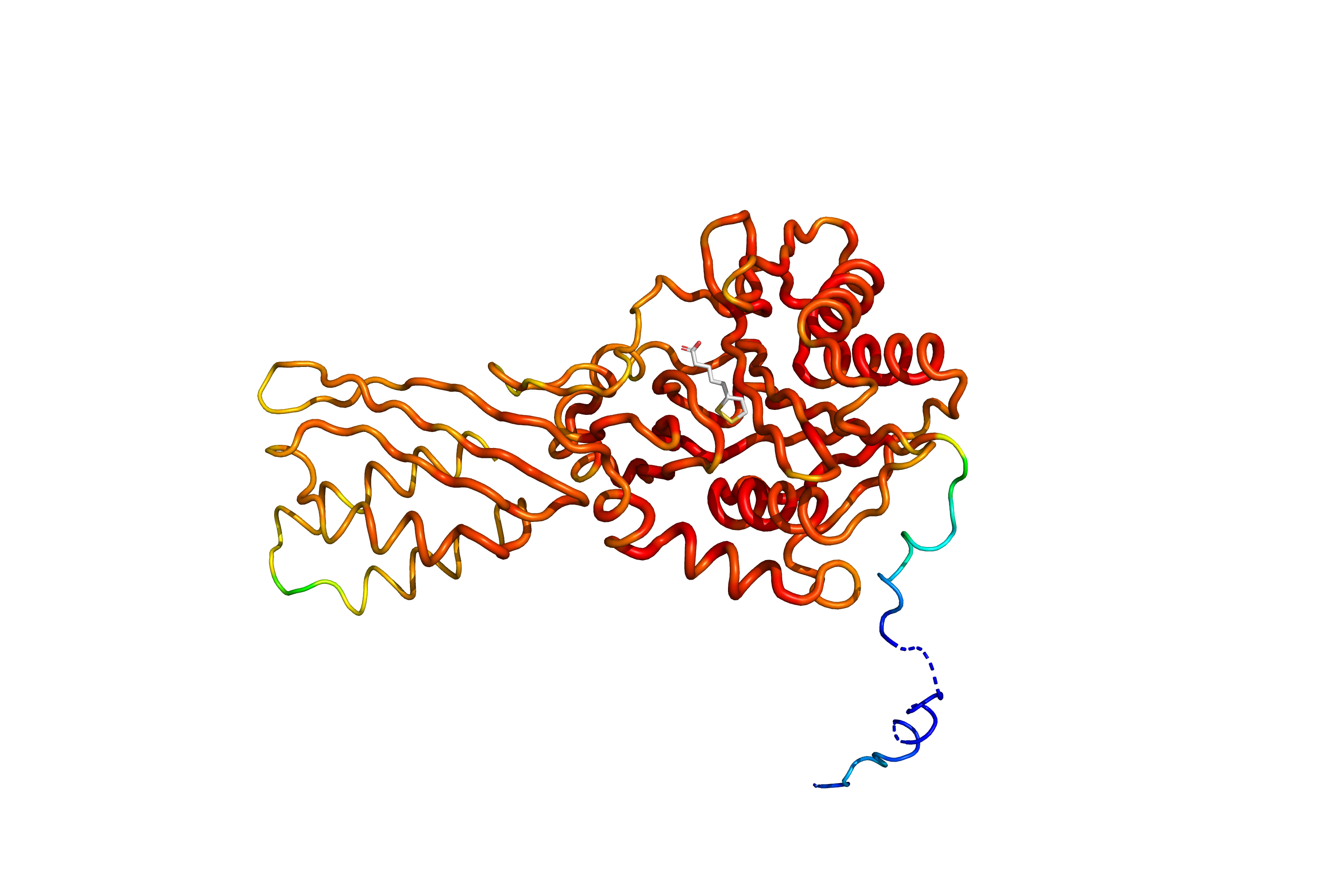}
    \caption{$B$ factor visualisation of RoseTTAFold All-Atom\cite{krishna2024generalized} prediction of LIPT1.}
    \label{fig:la_main_b_factor}
\end{figure}

\begin{figure}[!htb]
    \centering
    \includegraphics[width=.8\textwidth]{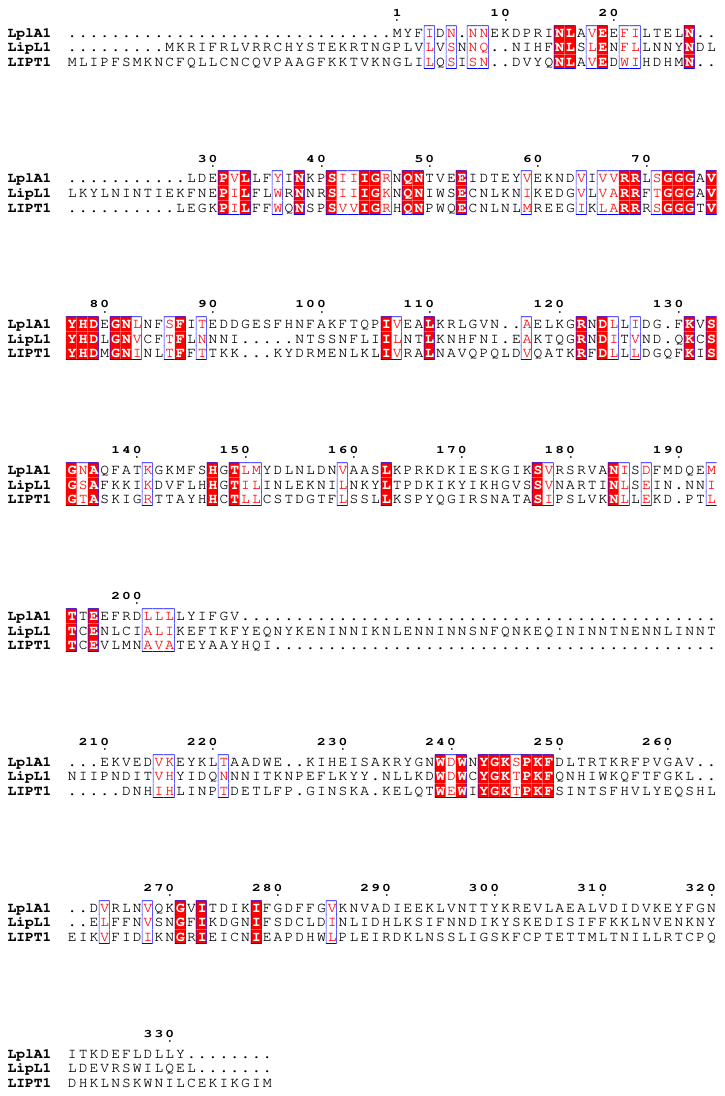}
    \caption{Sequence alignment of lplA1, LipL1 and LIPT1.}
    \label{fig:seq_align}
\end{figure}

\end{document}